\documentclass[acmsmall]{acmart}
\usepackage[flushleft]{threeparttable}
\usepackage[shortlabels]{enumitem}
\usepackage{xcolor}
\usepackage{subfigure}

\usepackage{amssymb}
\usepackage{pifont}
\usepackage{tcolorbox}
\usepackage{array}
\usepackage{tabularx}
\usepackage{geometry}
\usepackage{booktabs}
\usepackage{wrapfig}
\newcommand{\eat}[1]{}

\newcommand{\cmark}{\ding{51}}
\newcommand{\xmark}{\ding{55}}

\usepackage[ruled,vlined,linesnumbered]{algorithm2e}
\AtBeginDocument{%
  \providecommand\BibTeX{{%
    \normalfont B\kern-0.5em{\scshape i\kern-0.25em b}\kern-0.8em\TeX}}}

\setcopyright{acmcopyright}
\copyrightyear{2025}
\acmYear{2025}
\acmDOI{10.1145/1122445.1122456}

\acmJournal{JDIQ}
 \acmVolume{10}
 \acmNumber{12}
\acmArticle{1}
 \acmMonth{10}

\begin{document}

\title{A GenAI System for Improved FAIR Independent Biological Database Integration}


\author{Syed Nazmus Sakib}
\orcid{0009-0000-2653-2767}
\affiliation{%
  \institution{Department of Computer Science, University of Idaho}
 \streetaddress{875 Perimeter Drive}
 \city{Moscow}
 \state{Idaho}
 \country{USA}
 \postcode{83843}
}
\email{saki3064@vandals.uidaho.edu} 
 
\author{Kallol Naha}
\orcid{0000-0002-1815-234X}
\affiliation{%
  \institution{Department of Computer Science, University of Idaho}
 \streetaddress{875 Perimeter Drive}
 \city{Moscow}
 \state{Idaho}
 \country{USA}
 \postcode{83843}
}
\email{naha7197@vandals.uidaho.edu}

\author{Sajratul Y. Rubaiat}
\orcid{0009-0001-0367-644X}
\affiliation{%
  \institution{Department of Computer Science, University of Idaho}
 \streetaddress{875 Perimeter Drive}
 \city{Moscow}
 \state{Idaho}
 \country{USA}
 \postcode{83843}
}
\email{ruba3062@vandals.uidaho.edu}

\author{Hasan M. Jamil}
\authornote{Author for correspondence.}
\orcid{0000-0002-3124-3780}
\affiliation{%
  \institution{Department of Computer Science, University of Idaho}
 \streetaddress{875 Perimeter Drive}
 \city{Moscow}
 \state{Idaho}
 \country{USA}
 \postcode{83843}
}

\email{jamil@uidaho.edu}

%% \renewcommand{\shortauthors}{Trovato and Tobin, et al.}


\begin{abstract}

Life sciences research increasingly requires identifying, accessing, and effectively processing data from an ever-evolving array of information sources on the Linked Open Data (LOD) network. This dynamic landscape places a significant burden on researchers, as the quality of query responses depends heavily on the selection and semantic integration of data sources --processes that are often labor-intensive, error-prone, and costly. While the adoption of FAIR (Findable, Accessible, Interoperable, and Reusable) data principles has aimed to address these challenges, barriers to efficient and accurate scientific data processing persist.

In this paper, we introduce {\em FAIRBridge}, an experimental natural language-based query processing system designed to empower scientists to discover, access, and query biological databases, even when they are not FAIR-compliant. FAIRBridge harnesses the capabilities of AI to interpret query intents, map them to relevant databases described in scientific literature, and generate executable queries via intelligent resource access plans. The system also includes robust tools for mitigating low-quality query processing, ensuring high fidelity and responsiveness in the information delivered.

FAIRBridge's autonomous query processing framework enables users to explore alternative data sources, make informed choices at every step, and leverage community-driven crowd curation when needed. By providing a user-friendly, automated hypothesis-testing platform in natural English, FAIRBridge significantly enhances the integration and processing of scientific data, offering researchers a powerful new tool for advancing their inquiries.

\end{abstract}

\ccsdesc[500]{Information systems~Query languages}
\ccsdesc[500]{Information systems~Information integration}
\ccsdesc[300]{Information systems~Information retrieval}
\ccsdesc[500]{Human-centered computing~Human computer interaction (HCI)}
\ccsdesc[500]{Computing methodologies~Natural language processing}
\ccsdesc[300]{Computing methodologies~Knowledge representation and reasoning}
\ccsdesc[100]{Computing methodologies~Search methodologies}
\ccsdesc[300]{Applied computing~Computational biology}
\ccsdesc[300]{Applied computing~Bioinformatics}
\ccsdesc[500]{Applied computing~Life and medical sciences}
\keywords{FAIR, Natural Language Processing, Knowledge Representation, Information Retrieval, Computational Biology, Data Integration, Large Language Model, Deep Web Database, Scientific Inquiries}



\maketitle

\section{Introduction}
\label{intro}
A fundamental challenge in the modern life sciences is the effective utilization of the exponentially growing data distributed across the complex and expanding web of the Linked Open Data (LOD) network. While this interconnectedness holds immense potential for accelerating discovery, it simultaneously presents researchers with a daunting challenge: effectively leveraging this data requires navigating a complex ecosystem of diverse and often incompatible data sources. Even when guided by the principles of FAIR (Findable, Accessible, Interoperable, and Reusable) data \cite{wilkinson2016fair}, researchers encounter significant obstacles in their quest for relevant information. These obstacles stem not only from the sheer volume of data but also from its heterogeneity, its distribution across disparate platforms, and the varying degrees of adherence to FAIR principles \cite{feijoo2020evaluating, foster2022cuf, shanahan2022rethinking}.

Further compounding these challenges is the issue of data format heterogeneity. Data might be presented in raw HTML tables, requiring specialized parsing techniques, or it might be locked within non-downloadable formats, necessitating complex web scraping approaches. Even when data is accessible, integrating datasets from multiple sources often requires resolving schema inconsistencies and performing non-trivial data transformations. These technical hurdles can be particularly daunting for researchers lacking advanced computational expertise, effectively creating a barrier to entry for many who could otherwise benefit from the wealth of available data.

To illustrate practical difficulties, consider a researcher who seeks to investigate the relationship between histone genes and male infertility. Let us take this example query $Q$:

\begin{quote}
    \textit{Retrieve gene and protein information for all "H2A histone" genes from UniProt and associated infertility data from the Male Infertility Knowledgebase (MiKDB).}
\end{quote}

\noindent This seemingly straightforward query quickly devolves into a complex data wrangling exercise. Although resources like UniProt \cite{magrane2011uniprot} and MiKDB \cite{JosephM2021} offer valuable information, researchers must first be aware of their existence and then possess the technical skills to query them effectively. Retrieving all H2A histone genes from UniProt and cross-referencing them with infertility data in MiKDB is a non-trivial task. It requires navigating different database interfaces, understanding their specific query languages, and potentially performing manual data integration. Even when datasets adhere to FAIR principles, practical barriers remain that hinder their effective use. Consider the DisGeNET database \cite{PINER2021}, which provides both a SPARQL endpoint and a REST API to facilitate access to its RDF-based data. While these tools theoretically satisfy FAIR compliance, their practical utility demands a steep learning curve. Researchers must possess advanced knowledge of SPARQL for querying RDF data or the programming skills to integrate and manipulate data through REST APIs. This technical overhead can pose a substantial obstacle, especially for resource-constrained researchers who may lack the computational expertise or time to dedicate to mastering such tools.

Recent research, such as the BioNursery system \cite{JamilKG2024}, has begun to explore the use of large language models (LLMs) to facilitate biological hypothesis testing. BioNursery demonstrates the potential to partially automate complex queries. A more advanced approach, demonstrated in BioNursery, involves using a querying system called Needle \cite{JamilKG2024} to automatically cross-link databases such as MiKDB and DisGeNET. However, even this approach currently relies on manually constructed access protocols and schema matching schemes. We believe that the approach in Needle could benefit from an autonomous discovery of databases such as MiKDB and DisGeNET from PubMed abstracts and also from autonomous identification of access protocols and schema mapping information to facilitate end-to-end automation, which is the main focus of this article.

\subsection{Bird's-eye View of FAIRBridge}

To make such an information retrieval paradigm possible,  we developed a system that aims to improve the findability, accessibility, and interoperability of heterogeneous and unFAIR datasets toward ad hoc autonomous Natural Language Query (NLQ) processing over arbitrary deep-web databases. It functions as a search engine and facilitator that enhances the practical findability and accessibility of unFAIR data sources without altering the original data. By bridging the gap between user queries and hard-to-find datasets, FAIRBridge aids in the general principles of FAIR without directly modifying the FAIRness of the data itself. Our proposed system combines neural semantic search with automated data identification, extraction, and formatting techniques to locate relevant unFAIR data sources and transform them into more readily usable information. We use sentence transformers to build a vector database of PubMed abstracts \cite{reimers2019sentence}. We map NLQs into vector representations and identify relevant scientific literature in order to retrieve minimal information about query relevant interesting databases. We use GPT 4o as an agent and exploit a natural language processing (NLP) technique-based smart wrapper to isolate necessary schema information to develop access protocols and schema heterogeneity resolution strategies. Finally, extraction techniques are used using this information to automatically retrieve targeted information. For the purpose of this article, we consider the ability to interrogate remote databases in FAIRBridge addressed Reusability. However, the issue of Reusability demands a more extensive treatment in its own right \cite{CamposRABFSG20, PatraSDWMWZRWY20, CousijnHKM22, Grant22}.
% In accordance with FAIRshake criteria \cite{clarke2019fairshake}, we develop practical evaluation metrics to quantitatively demonstrate its findability capabilities.

%%%%%%%%%%%% qualitative analysis %%%%%%%%%%%%
To evaluate the effectiveness of FAIRBridge in facilitating the discovery and access of unFAIR biological data sources, we conducted a comparative analysis against several existing information retrieval (IR) systems. Table \ref{tab:ir_systems} summarizes the features of various IR systems, including Google Dataset Search, DataCite, re3data, Zenodo, Figshare, Dryad, PANGAEA, ICPSR, and PubMed, in comparison to FAIRBridge.

% Comparative Analysis of FAIRBridge with Existing IR Systems

\begin{table*}[!htbp]
    \centering
    \small
    \caption{Comparison of IR Systems}
    \label{tab:ir_systems}
    \begin{tabular}{|p{3.25cm}|*{9}{>{\centering\arraybackslash}p{0.8cm}|}}
      \hline
        \textbf{Dataset Search Tool}
          & \rotatebox{90}{\shortstack[c]{\textbf{NL}\\\textbf{Query}}}
          & \rotatebox{90}{\shortstack[c]{\textbf{Cross-Portal}\\\textbf{Search}}}
          & \rotatebox{90}{\shortstack[c]{\textbf{Custom Data}\\\textbf{Formatting}}}
          & \rotatebox{90}{\shortstack[c]{\textbf{Data}\\\textbf{Presentation}}}
          & \rotatebox{90}{\shortstack[c]{\textbf{Metadata}\\\textbf{Display}}}
          & \rotatebox{90}{\shortstack[c]{\textbf{Web}\\\textbf{Crawling}}}
          & \rotatebox{90}{\shortstack[c]{\textbf{Form}\\\textbf{Processing}}}
          & \rotatebox{90}{\shortstack[c]{\textbf{Unfair Hand-}\\\textbf{ling Data}}}
          & \rotatebox{90}{\shortstack[c]{\textbf{Customi-}\\\textbf{zable Output}}} \\
        \hline
        
        Google Dataset Search \cite{47845}
          & Limited & \cmark & \xmark & \xmark & \cmark & \cmark & \xmark & \xmark & \xmark \\
        \hline
        DataCite \cite{datacite}
          & \xmark & \cmark & \xmark & \xmark & \cmark & \xmark & \xmark & \xmark & \cmark \\
        \hline
        re3data \cite{re3data}
          & Limited & \cmark & \xmark & \xmark & \cmark & \xmark & \xmark & \xmark & \xmark \\
        \hline
        Zenodo \cite{zenodo}
          & \xmark & \xmark & \cmark & \cmark & \cmark & \xmark & \xmark & \xmark & \cmark \\
        \hline
        Figshare \cite{figshare}
          & \xmark & \xmark & \cmark & \cmark & \cmark & \xmark & \xmark & \xmark & \cmark \\
        \hline
        Dryad \cite{dryad}
          & \xmark & \xmark & \xmark & \cmark & \cmark & \xmark & \xmark & \xmark & \cmark \\
        \hline
        PANGAEA \cite{pangaea}
          & \cmark & \cmark & \xmark & \cmark & \cmark & \xmark & \xmark & \xmark & \cmark \\
        \hline
        ICPSR \cite{icpsr}
          & Limited & \xmark & \xmark & \cmark & \cmark & \xmark & \xmark & \xmark & \cmark \\
        \hline
        Pubmed
          & Limited & \xmark & \xmark & \xmark & \cmark & \xmark & \xmark & \xmark & \cmark \\
        \hline
        FAIRBridge
          & \cmark & \cmark & \cmark & \cmark & \cmark & \cmark & \cmark & \cmark & \cmark \\
        \hline
    \end{tabular}
\end{table*}

% Cross-Portal Search & Query Support:
DataCite and re3data enable cross-portal searches but rely on basic keyword-based queries. Google Dataset Search and PANGAEA offer limited natural language support but still depend on keyword matching. In contrast, FAIRBridge provides full natural language query support using advanced language models, allowing retrieval without precise keywords.
% Data Presentation & Customization: 
While platforms like Zenodo, Figshare, and Dryad host datasets, they are limited to their own repositories and lack output customization. FAIRBridge retrieves data across platforms and allows flexible dataset reshaping based on user-defined parameters, catering to specific research needs.

% Handling of UnFAIR Data & Form Input Processing: 
Existing IR systems do not explicitly address unFAIR data retrieval or form-based data access. FAIRBridge overcomes this by using a smart data wrapper to extract key information from web resources, including those requiring form submissions, enabling access without modifying the original infrastructure.

% Web Crawling & Knowledge Base Integration: 
Google Dataset Search indexes datasets via schema.org markup but does not process actual data. DataCite and re3data depend on repository submissions. FAIRBridge combines web crawling with a vector database built from PubMed abstracts, enhancing semantic retrieval. Its integrated knowledge base further improves query understanding, unlike systems relying solely on metadata schemas.

% Custom Data Formatting & Output: 
Zenodo and Figshare support custom formatting but only within their repositories. FAIRBridge extends this capability across platforms, reshaping datasets based on user criteria—offering a level of flexibility absent in other IR tools.

\subsection{FAIR Compliance Gap}

The FAIRBridge tool addresses a critical gap in the current landscape of scientific data retrieval and utilization. In the past, researchers have located and accessed datasets manually through time-consuming and ineffective methods like Google searches and PubMed enquiries. The majority of PubMed searches rely on mesh headings, which makes it difficult to locate the most pertinent papers \cite{pubmed_search}. The data retrieval process is further complicated by the fact that the data found using these techniques are frequently hosted on sites that do not follow schema.org standards or FAIR principles. These platforms potentially need user input, which would further complicate data access.
By automating data retrieval and discovery, FAIRBridge transforms this process and improves the accessibility and interoperability of scientific datasets. Unlike other FAIR data sites that are constrained by the data provided by users, FAIRBridge goes beyond by enabling the detection of unFAIR data without demanding that the data itself comply to FAIR criteria. By removing the obstacles in the way of achieving FAIR data, this tool offers researchers a quick and easy answer. FAIRBridge ensures that data is not only easier to locate but also more usable by utilising advanced semantic search capabilities, cross-portal integration, and direct data retrieval. This frees up researchers to concentrate on their scientific enquiries rather than the tiresome chore of data hunting.

\subsection{FAIRBridge Design Goals}

The primary design goal of FAIRBridge is to empower researchers to interact with the vast and complex landscape of biological data using intuitive natural language queries. The system is designed to be powerful and user-friendly, automating complex data retrieval tasks while providing transparency and control to the user. Key contributions include:

\begin{itemize}
    \item \textbf{Autonomous Database Discovery:} FAIRBridge intelligently identifies potentially relevant databases from scientific literature and other sources.
    \item \textbf{Automated access protocol generation:} The system automatically determines how to access and query each database, handling diverse interfaces and formats.
    \item \textbf{Schema heterogeneity resolution:} FAIRBridge employs advanced techniques to resolve inconsistencies between different data schemas, enabling integration.
    \item \textbf{Natural Language Query Interface:} Users can interact with the system using intuitive natural language, eliminating the need for specialized query languages.
    \item \textbf{Enhanced Data Interoperability:}

    Interoperability is ensured by enabling SQL-like operations such as JOIN, LIKE, and filtering to merge and harmonize datasets from multiple sources. This module bridges schema and format heterogeneity, and present researchers with unified, usable data outputs.
    \item \textbf{Human-in-the-loop capabilities:} FAIRBridge offers optional human-in-the-loop functionality, allowing users to guide the system and refine results when needed.
    \item \textbf{Rich and Evolving Database of Process Descriptions:} Establishing a robust database of access schemas, or process descriptions, that encapsulate the method of interacting with diverse resources. This repository becomes richer with increased user engagement, as new interactions and workflows contribute to a growing knowledge base. This enhancement ensures faster, more accurate resource discovery and minimizes redundancy over time.
\end{itemize}

By achieving these design goals, FAIRBridge aims to democratize access to biological data, empowering researchers of all levels of computational skills to explore complex scientific questions and accelerate the pace of discovery. The following sections will explore the architecture and implementation of FAIRBridge, followed by a detailed evaluation of its performance and a discussion of its implications for the future of data-driven life sciences research.

\section{Related Works}
\label{related}

Machines' capacity to discover and analyze data autonomously, accurately and efficiently across diverse internet sources can be aided with FAIR compliance.
In order to make search and retrieval easier using FAIRness, significant strides have been made in many application areas with adherence to Wilkinson et al.'s \cite{wilkinson2018design} FAIRness model and their proposed metrics for digital resources based on community-defined Maturity Indicators (MIs) and Compliance Tests. Among them, Clarke et al. \cite{clarke2019fairshake} introduce a web-server application called the FAIRshake, a database, an API, a browser extension, a bookmarklet, and FAIR analytics modules. FAIRshake also provides FAIR metrics and rubrics, which are questions and collections of questions to assess a digital object's FAIRness. Devaraju et al. \cite{devaraju2020fairsfair} present 15 metrics to measure how FAIR research data objects are, based on existing work and indicators from the RDA FAIR Data Maturity Model. The purpose of the metrics is to develop practical solutions to facilitate the application of the FAIR principles throughout the life cycle of the research data.

% CIKM UPDATE
For the purpose of quantitative exposition of what FAIRness offers, Devaraju et al. \cite{devaraju2021automated} developed a programmatic approach to assessing the FAIRness of data, and using their ``F-UJI'' system, they measured the FAIRness of about 2,500 data objects from five diverse domains and summarized their findings in the form of recommendations for improvement. E. Amdouni et al. \cite{amdouni2022faire} too introduce 61 questions for FAIRness assessment, with 80\% based on resource metadata about digital objects'  licenses, examples, provenance, and documentation attributes.

% \eat{
% Efforts in the geosciences \cite{stall2017enabling}  and biological sciences \cite{lannom2020fair} show how dataset features and variables may be cataloged using FAIR, allowing for interoperable search services across hundreds of databases.
% }

% A growing number of research have concentrated on methods for searching unstructured text contents of internet documents, as opposed to the structured databases of metadata that were the initial emphasis of many IR systems. Several recent systems have explored improved relevance ranking algorithms for retrieving documents that match users' queries from large internet databases. Kumar et al. \cite{kumar2015identifying} developed a novel algorithm, Latent Semantic Manifold (LSM), which outperformed other contemporary algorithms in identifying semantic topics in high-dimensional web data. Landauer et al. \cite{Landauer2008LatentSA} proposed a method for automatic indexing and retrieval using singular-value decomposition, which showed promising results in initial tests.  Khin et al. \cite{WaiKhin2018QueryCB} focuses on query classification, using a Web Query Classification Algorithm to categorize queries and retrieve relevant documents. 

Traditional IR systems often relied on lexical matching techniques, such as Boolean models, vector space models, and probabilistic models \cite{goyal2011query}. These approaches match query terms with document terms, but they may struggle with vocabulary mismatch, synonymy, and polysemy issues, leading to sub-optimal retrieval performance \cite{BernsteinMR11}.

Cooperative Query Answering Database Systems (CQADS) aim to provide cooperative and informative responses to user queries by relaxing query conditions or suggesting alternative queries \cite{edgebase}. However, CQADS lacks precision in context awareness across multiple data sources, and does not provide similar datasets along with the requested ones. 

With the advent of deep learning and natural language processing (NLP) techniques, neural IR models have gained popularity for their ability to capture semantic relationships and contextual information. Representation-based models, such as DSSM \cite{huang2013learning} and CDSSM \cite{shen2014latent}, learn dense vector representations of queries and documents, which are then compared using similarity measures like cosine similarity or dot product. Interaction-based models, such as DRMM \cite{guo2016deep} and K-NRM \cite{xiong2017end}, focus on modeling the interactions between query and document terms, capturing fine-grained relevance signals.

The development of NLP techniques and language models has significantly enhanced the power of information retrieval systems \cite{QingyaoIR}. With the vast amounts of unstructured text data available on the internet and in enterprise document collections, there is a need for systems that can interpret and match user queries that are posed in natural language.

More recently, transformer-based language models like BERT \cite{devlin2018bert} and its variants have been successfully applied to IR tasks, leveraging their ability to capture contextual information and long-range dependencies. Models like MonoBERT, DuoBERT \cite{nogueira2019multi}, and ColBERT \cite{khattab2020colbert} have achieved state-of-the-art performance on benchmark IR datasets like MS MARCO \cite{nguyen2016ms} and TREC \cite{voorhees2005trec}.

% While these neural IR models have shown promising results, ensuring the adherence to FAIR principles remains a challenge, especially for biomedical literature. Several studies have explored the use of knowledge graphs and ontologies to enhance the findability, accessibility, and interoperability of biomedical information. BioSyn [13] utilizes biomedical ontologies and knowledge graphs to improve the retrieval and ranking of biomedical abstracts. Similarly, the BioLitCry system [14] employs domain-specific knowledge bases and semantic annotations to enhance the retrieval and exploration of biomedical literature.

% Despite these efforts, ensuring the reusability of retrieved biomedical abstracts remains a significant challenge, as it often requires additional metadata, provenance information, standardized formats and an end to end system. Approaches like FAIR Data Point [15] and FAIR Datasets [16] have been proposed to facilitate the reusability of scientific data, but their applicability to biomedical literature retrieval has not been extensively explored.
% NLP tasks including question answering, phrase semantic similarity assessment, and document rating have demonstrated notable effectiveness with recent neural language models, such as BERT, that are pre-trained on huge text corpora \cite{devlin2018bert}. 

While prior work has made advances in individual areas like defining FAIR metrics \cite{wilkinson2018design, wilkinson2016fair}, building FAIR repositories for specific domains \cite{lannom2020fair}, and enhancing information retrieval from text \cite{devlin2018bert, guo2016deep}, to our knowledge no previous system has provided an integrated, fully-automated approach to improve the findability, accessibility and interoperability of unFAIR biological data sources across the literature. Our FAIRBridge system is a novel solution that combines a neural semantic search, web mining, and data extraction techniques in a flexible, end-to-end workflow.

% For information retrieval purposes, these pre-trained models can be tweaked and modified to fit application needs. To capture underlying semantic meanings, models can also be trained to learn latent vector representations of query and document terms and sentences. This allows systems to match relevant documents even when keyword terms in the query and results differ \cite{guo2016deep}.
 
% As users increasingly expect search engines to comprehend and respond to conversational NLQs, advancements in semantic parsing from NLP research are being utilized in commercial systems as well \cite{min2018efficient}. 

% Information retrieval systems now have additional ability to make sense of verbose or imprecise query input and match relevant content in huge unstructured text corpora with the underlying information needs thanks to the combination of neural language models and NLP. In order to deliver more naturalistic search experiences, these strategies will keep developing.

% \section{Bird's Eye View of FAIRBridge}
% \label{FBS}

\section{Features of FAIRBridge}
\label{tour}

\begin{figure}[ht!]
  \centering
 \includegraphics[width=.8\textwidth,keepaspectratio]{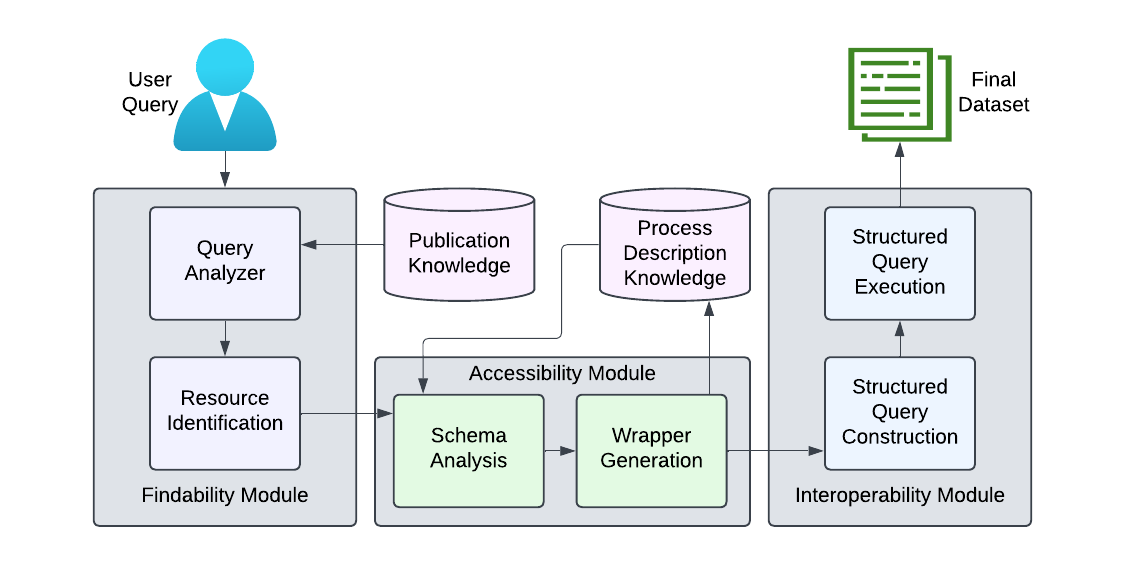}
 \caption{Functional Components of FAIRBridge}
  \label{fig:arch}
\end{figure}

Ahead of a more technical discussion, it is perhaps useful to present an overall idea about the FAIRBridge system from the perspective of the end users using a simple illustrative example. The intent is to elucidate the purpose of the FAIRBridge design and support that scientists can expect from it. While Fig \ref{fig:arch} depicts the functional components of FAIRBridge that make up its architecture, Figs \ref{fig:fe1} and \ref{fig:fe2} depict its front-end user interfaces in action. Fig \ref{fig:fe1} shows the landing page of FAIRBridge where users initiate their smart querying exercise using natural English paragraphs. They will initiate a query by asking a question in the top text box. The format is free-form and users are allowed to ask the query in their preferred way. 
% There is no limit on the number of characters, words, lines or paragraphs.

They are also allowed to add additional information, or knowledge, the user believes could help expedite, improve or narrow down the search for a response using the text box just below. FAIRBridge will use this additional information in its formulation of the query response. Users can choose between two querying options -- a) full automatic computation, or b) guided or human-in-the-loop computation. The current edition of FAIRBRidge also includes example queries that the new users could try from a drop-down list at the top. The current edition does not allow storage. Our plan is to allow users to have personalized space to enable community interaction in the near future.

\subsection{Automated Scientific Inquiries}

\begin{figure}[ht!]
\centering
\subfigure[FAIRBridge landing page.]{
    \includegraphics[width=.65\textwidth,keepaspectratio]{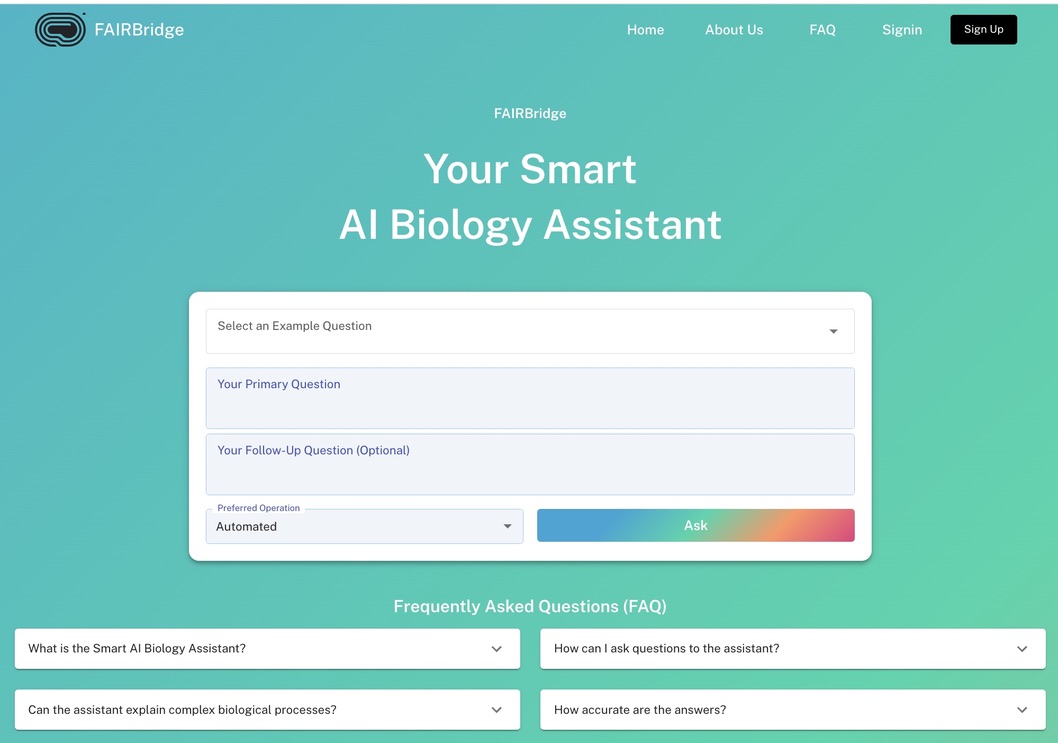}
    \label{fig:fe1}
}
\subfigure[User interaction on FAIRBridge.]{
 \includegraphics[width=0.85\textwidth,keepaspectratio]{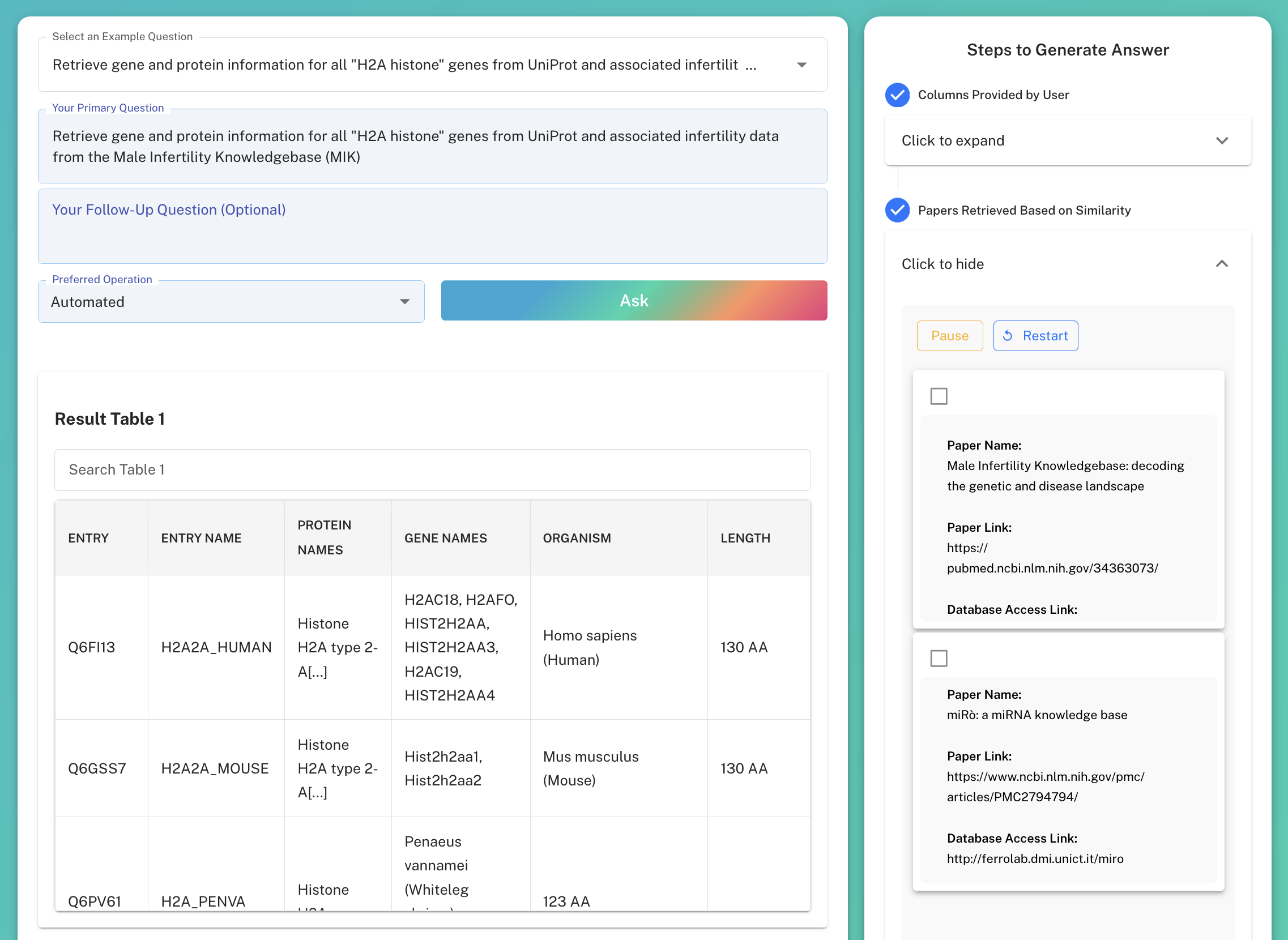}
  \label{fig:fe2}
}
\caption{FAIRBridge user interface.}
\label{fig:userinterface}
\end{figure}

Once a query is posed, FAIRBridge analyzes the query intent and formulates a query processing plan as its default automated execution mode. The plan includes identification of the needed resources, generating resource access plans, database schema heterogeneity resolution for data integration, and query processing plan. Since a query may potentially be computed in multiple different ways, these choices result in forest of possibilities, from which the system chooses the most appropriate sequence. In other words, at every step, FAIRBridge ranks all the available choices, selects the top ranked options and proceeds to the next step in a similar fashion. Therefore, the ultimate sequence is a branch of a tree of possible options. During the execution of the query, this forest of choices is shown on the right hand side panel (Fig \ref{fig:fe2}) as a reference, and displays the result at the bottom panel. Since the FAIRBridge conceptual model uses a relational model of data, all views of data are tabular in the sense of relational data model. Fig \ref{fig:fe2} shows the processing and the result of the example query $Q$.
% \begin{quote}
%     Retrieve gene and protein information for all "H2A histone" genes from UniProt and associated infertility data from the Male Infertility Knowledgebase (MiKDB)
% \end{quote}
Note that this query requires retrieving all Histone genes in UniProt in a local table, and then performing a natural join operation with the genes collected in a similar way, again in a local table from the MiKDB database over a possible heterogeneous scheme once a suitable access plan is generated, all without any user assistance. While the technical details are presented in the sections ahead, it is expository to appreciate that the challenges FAIRBridge addresses are quite significant.

% \begin{figure}[ht!]
% \centering

%  \includegraphics[width=0.85\textwidth,keepaspectratio]{from_Kallol_01/multiple_auto.png}

% \caption{User interaction on FAIRBridge.}
% \label{fig:fe2}
% \end{figure}

FAIRBridge uses the capabilities of GPT 4o to identify the resources from PubMed abstracts or a local vector database, and then analyzing them to discover the possible scheme of each database. ChatGPT has been trained to map the access information as a Needle resource description for onward processing. An essential step then is to generate wrappers to abstract the lower level access details and model the data into tables. Once all the resources are processed and the corresponding Needle descriptions are generated, FAIRBridge generates a sequence of BioFlow \cite{MouJ20} queries to process the query plan.

\subsection{Guided Scientific Inquiries}

FAIRBridge is designed to handle diverse and complex scientific queries, using advanced technologies such as ChatGPT. However, due to the inherent complexity of the FAIRBridge workflow, there may be instances where the tool requires additional user guidance to achieve optimal results. This can occur in scenarios such as identifying a suitable resource from the literature, determining the correct schema and access methods for a database, managing session timeouts or broken processes, or refining wrapper generation. Additionally, constructing a semantically meaningful query processing plan that fully reflects the user's intent is a nuanced task that occasionally benefits from user input.

To address these situations, FAIRBridge provides a user-guided querying approach that empowers users to participate in the decision-making process. This approach simplifies tasks such as selecting the right databases or tools, reviewing and refining schema information, modifying BioFlow queries, or choosing the computational logic for workflows. FAIRBridge provides users with straightforward choices, empowering them to make informed decisions without needing to navigate the system’s technical complexities. It also facilitates collaboration by offering the capability to involve the broader community in resolving complex tasks when needed.

To further enhance usability, FAIRBridge introduces a "Guided Execution" mode, allowing users to steer computations step by step. This mode presents relevant options for each stage of the process in an interactive panel, enabling users to select the most appropriate choices. Users can also edit schema descriptions, wrappers, and query logic or involve the community for assistance. This interactive framework ensures that users remain in control, guiding the semantic interpretation of their queries, while FAIRBridge efficiently manages the underlying technical operations. By doing so, FAIRBridge bridges the gap between user intent and application logic, allowing researchers to focus on their scientific inquiries without needing to engage with the complexities of coding or application workflows.

\section{System Architecture}
\label{sec:system}

FAIRBridge is deployed as a client-server application, featuring a React-based \cite{react} frontend and a Python Flask \cite{flask} backend. This section presents the software stack, hardware infrastructure, and key design decisions underpinning the platform.

\subsection{User Interface}

The frontend is built with React, chosen for its component-based approach and strong community support. Material UI \cite{material-ui} and Ant Design \cite{ant-design} provide ready-to-use UI components, ensuring a consistent, modern look. Development and build processes rely on Vite \cite{vite} for speed and efficiency, with Node.js managing dependencies and scripts.

\subsection{Backend}

The lightweight Flask backend handles core logic such as natural language processing, data retrieval, and interaction with the frontend via RESTful \cite{fielding2000} endpoints. Query interpretation relies on GPT-4o \cite{gpt-4o}, accessed via the OpenAI API and integrated with langchain \cite{langchain}. A “Smart Wrapper” component utilizes Beautiful Soup \cite{richardson2007beautiful} for HTML/XML parsing and Selenium \cite{selenium} for dynamic web pages. Schema matching techniques, inspired by CUPID \cite{madhavan2001generic} and COMA \cite{do2002coma}, unify data from heterogeneous deep-web databases.

Data integration is managed with pandas for manipulation and DuckDB \cite{DBLP:conf/sigmod/RaasveldtM19} for in-memory analytical queries on DataFrames. A Chroma vector database \cite{chroma} stores and indexes embeddings of scientific literature, enabling semantic search and quick source identification.

\subsection{Hardware}

FAIRBridge runs on a Rocky Linux 8 server with 16 CPU cores, 32 GB of RAM, and a 256 GB NVMe SSD. This setup efficiently handles concurrent queries and large datasets, consuming around 350 MB of memory per NLP query. The combination of carefully selected libraries, frameworks, and robust hardware allows FAIRBridge to rapidly retrieve, integrate, and present biological data from diverse sources.
% Table \ref{tab:libraries} provides a concise summary of these key technologies.

% have to mention about langchain. 

% \begin{table}[h]
% \centering
% \caption{Key Libraries and Frameworks Used in FAIRBridge}
% \label{tab:libraries}
% \begin{tabular}{ll}
% \toprule
% Library/Framework & Description \\
% \midrule
% React & JavaScript library for building user interfaces \\
% Material UI & Comprehensive suite of pre-built UI components \\
% Ant Design & Enterprise-class UI design language and React UI library \\
% Vite & Fast and efficient build tool supporting TypeScript \\
% Node.js & Server-side runtime environment for JavaScript \\
% Flask & Python web framework \\
% Beautiful Soup & Python library for parsing HTML and XML \\
% Selenium & Python library for web browser automation \\
% pandas & Python library for data manipulation and analysis \\
% langchain & Python library for building applications with LLMs\\
% DuckDB & In-process SQL OLAP database management system \\
% Chroma & Vector database for storing and retrieving embeddings \\
% GPT-4o & Large language model from OpenAI \\
% OpenAI API & API for interacting with OpenAI models \\
% \bottomrule
% \end{tabular}
% \end{table}

\section{Implementation Strategies of FAIRBridge}
\label{arch}

The architecture of FAIRBridge depicted in Fig \ref{fig:arch}, and its functional intent discussed in Sec \ref{tour} anticipate several technical solutions that we developed and discussed in this section. In this section, we will introduce the FAIRBridge query processing algorithm that leverages these technologies to capture the vision presented in Sec \ref{tour} that the users will experience.

% \subsection{Implementation of FAIRBridge}

% The FAIRBridge system exploits several tools for it's implementation using the Python Flask framework. We use the open source Chroma DB as our vector database. Its primary purpose is to facilitate the storage and retrieval of vector embeddings along with relevant metadata. In order to explore, retrieve and  analyze web contents, we use Selenium. We use GPT 4o as an agent to handle language processing tasks, enhancing query processing and data retrieval. These technologies are used to implement the three main modules of FAIRBridge as shown in Fig. \ref{fig:arch} and briefly highlighted in the sections below.

\subsection{Findability Module}
This module is implemented to ensure findability. Our initial initiative involves establishing a publication database comprised of 567 data points. The database is extracted from "The Journal of Biological Databases and Curation" spanning 2009 to 2023. Each data point includes document metadata—title, abstract, database access link, and publication year. To enable efficient retrieval, we utilize the "text-embedding-ada-002" embedding model for vectorization. These embeddings capture the semantic content of paper-abstracts, complemented by stored metadata for each data point. Generating vector representations of queries using sentence encoders allows integration with downstream vector similarity-based algorithms \cite{devlin2018bert}. This module performs two functionalities:
%\cite{reimers2019sentence} 
% \begin{algorithm}
% \caption{Query Reformatter}
% \label{alg:query_format}
% \LinesNumbered
% \KwIn{Query $Q$}
% \KwOut{Optimized search query $Qs$, Metadata $M$}
% Parse $Q$ to identify key parameters\;
% Generate optimized search query $Qs$\;
% Construct metadata $M$ (data format, source, description, etc.)\;
% \Return $Qs, M$\;
% \end{algorithm}

% \begin{algorithm}
% \caption{Query Processor}
% \label{alg:query_processor}
% \LinesNumbered
% \KwIn{User Query $q$}
% \KwOut{Retrieval Query $\tilde{q}$, Expanded Queries $Q_e$, Metadata $M$}
% Parse $q$ to identify key parameters\;
% Generate optimized search query $\tilde{q}$\;
% Construct metadata $M$ (data format, source, description, etc.)\;
% Convert $\tilde{q}$ to vector representation $v_{\tilde{q}}$\;
% Compute similarity scores between $v_{\tilde{q}}$ and vectors in the database\;
% Retrieve top-$n$ similar contexts $C = \{c_1, c_2, \dots, c_n\}$\;

% Generate expanded query $\hat{q}_i$ using RAG based on $\tilde{q}$ and $C$\;

% Form the expanded query set $Q_e = \{\hat{q}_1, \hat{q}_2, \dots, \hat{q}_k\}$\;
% \Return $\tilde{q}$, $Q_e$, $M$\;
% \end{algorithm}

\subsubsection{\textbf{Query Analyzer:}}
To accommodate user queries, which can be diverse in structure and content, we implement a query processor function to transform free-form textual queries into optimized search representations. Alg. \ref{alg:query_processor} presents the function of query formation and expansion where we transform user inputs into a standardized format and expand the queries. Using GPT 4o, we generate a JSON representation of the query containing key information. For instance, we take the example query $Q$ from Section \ref{intro}.

We employ the query expansion technique to expand the retrieval query $\tilde{q}$ \cite{koo2024optimizing}. It is converted into a vector representation $v_q$. This vectorization enables the computation of semantic similarities between the query and entries in the vector database. The top-n similar contexts $C = \{c_1, c_2, ..., c_n\}$ are retrieved based on the computed similarity scores. These contexts provide additional semantic information that can be used to enrich the query. Using Retrieval-Augmented Generation (RAG), the expanded query set $Q_e = \{Q_1, Q_2, ..., Q_k\}$ is formed, which will be used in subsequent modules. Here is the query set $Q_e$ from $\tilde{q}$:

\begin{enumerate}
    \item \textit{"Retrieve gene mutations associated with H2A histone proteins from UniProt and MiKDB."}
    \item \textit{"Access datasets on infertility factors linked to H2A histone genes."}
    \item \textit{"Retrieve protein length, organism details, and gene names for H2A histones from UniProt."}
    \item \textit{"Get data on reproductive genetics focusing on teratozoospermia."}
    \item \textit{"Extract genetic information on male infertility and histone-related disorders."}
\end{enumerate}

A set of keywords $K$ is generated for the PubMed search function, containing relevant terms that represents the query.

\begin{table}[htbp!]
\centering
\small
  \caption{Data Sources for the Retrieval Query}
  \label{tab:data_sources}
  \centering
  \begin{tabular}{|p{2cm}|p{4cm}|p{4cm}|}
    \hline
    \textbf{Field} & \textbf{Source 1} & \textbf{Source 2} \\
    \hline
    \textit{retrieval query} & H2A histone & H2A histone \\
    \hline
    \textit{data source name} & Male Infertility Knowledgebase (MiKDB) & UniProt \\
    \hline
    \textit{data link} & \href{http://mik.bicnirrh.res.in/}{mik.bicnirrh.res.in} & \href{https://www.uniprot.org/}{uniprot.org} \\
    \hline
    \textit{paper title} & Male Infertility Knowledgebase: decoding the genetic and disease landscape & The UniProtKB guide to the human proteome \\
    \hline
  \end{tabular}
\end{table}

\subsubsection{\textbf{PubMed Search:}}
    The PubMed search module in Algorithm \ref{alg:pubmed_search} leverages a combinatorial search technique to overcome limitations in handling natural language queries. 
    PubMed API was designed for keyword-based search rather than complex textual queries \cite{biopython_pubmed, jin2023pubmed}. To improve recall, the algorithm relies on a set of keywords $K$, derived from the query $Q$, rather than using the query itself. This allows for automatic reformulation into a series of keyword-based search queries. Specifically, the extracted keywords are input into an iterative engine that generates combinations of decreasing length. First, all keywords are combined into a conjunction. If this returns no results, the search drops one term and retries with $n-1$ keywords. The process repeats, eliminating one keyword in each iteration until a result is found or until the combination reaches the minimum threshold of $L$ keywords.

\subsubsection{\textbf{Resource Identification:}}
As presented in Alg. \ref{alg:database_search}, after obtaining the retrieval query \( \tilde{q} \) and the expanded query set \( Q_e \) from the Query Processor function, we proceed with the database search. Each expanded query \( \hat{q}_i \) in \( Q_e \) is transformed into its vector representation \( v^{q_i} \), placing it within the same high-dimensional vector space as the database documents.

For each expanded query \( \hat{q}_i \), we compute the cosine similarity between \( v^{q_i} \) and the vectors representing the documents \( v^{d_j} \) in the database \( \Delta \). Cosine similarity measures the angular similarity between vectors, quantifying the closeness between each expanded query vector and the document vectors. Based on these similarity scores, we retrieve the top-\( n \) documents \( D_i = \{ d_{i1}, d_{i2}, \dots, d_{in} \} \) that are most relevant to \( \hat{q}_i \).

The documents retrieved from all the expanded queries and PubMed documents are then merged into a single collection \( D \), and duplicates are removed to ensure uniqueness. We then convert the retrieval query \( \tilde{q} \) into its vector representation \( v^{\tilde{q}} \). To rank the merged documents, we compute the cosine similarity between \( v^{\tilde{q}} \) and each document vector \( v^{d_j} \) in \( D \).

% \[
% \text{sim}_{\text{final}}(v^{\tilde{q}}, v^{d_j}) = \text{cosine\_similarity}(v^{\tilde{q}}, v^{d_j})
% \]

These similarity scores quantify the closeness between the retrieval query vector and the document vectors, allowing us to rank the documents based on their relevance to \( \tilde{q} \). Higher similarity scores indicate greater relevance to the query. The documents in \( D \) are then ranked according to their \( \text{sim}_{\text{final}} \) scores. The top-ranked data points are subsequently presented as the search results. This process ensures that the final results are highly relevant to the user's refined intent expressed in the retrieval query \( \tilde{q} \). The top-ranked database sources are then passed to the Accessibility Module for further analysis and data extraction.

In the next step, the system identifies the necessary data sources for executing the user query by leveraging structured prompt techniques semantic reasoning and the top-ranked database sources. This phase produces a JSON representation of the query, specifying the relevant data sources, their access links, and corresponding metadata. This output serves as a precursor to retrieving process descriptions and executing subsequent operations.

The resource identification process takes the user query $Q$ and the top-ranked data sources $\Delta$ to produce a structured output containing details of the required data sources that match the user query. This structured output provides essential metadata such as retrieval queries, data source names, data access links, and corresponding research publications.

The function leverages GPT-4o with structured prompting to ensure accurate identification of relevant data sources. The function utilizes a pre-defined schema to enforce consistency and robustness in the outputs.

The structured prompt used in this step includes the following components:
\begin{itemize}
    \item     Query Context: The user query is analyzed to extract intent and contextual information.
    \item     Metadata Integration: Metadata from the database search results is incorporated into the prompt to refine the search for relevant data sources.
    \item   Response Schema: A predefined schema ensures that the output conforms to a specific format, aiding downstream processes.
\end{itemize}

Table \ref{tab:data_sources} shows the results of the resource identification stage. The output of this stage is a structured JSON representation that details the relevant data sources, their access links, and associated metadata. This representation provides inputs for subsequent operations, including retrieving process descriptions and executing the BioFlow query which will be discussed later.

Each data source is described with a retrieval query, indicating the term or condition to search for, a data source name specifying the resource name, a data link providing the URL for direct access, and a paper title offering metadata that contextualizes the data source. This structured output ensures that the system has all the necessary information to access, retrieve, and process the required data efficiently.

\begin{table}[htbp!]
\begin{minipage}[t]{0.48\textwidth}
\begin{algorithm}[H]
\caption{Query Processor}
\label{alg:query_processor}
\LinesNumbered
\KwIn{User Query $Q$}
\KwOut{Retrieval Query $\tilde{q}$, Expanded Queries $Q_e$, Keywords $K$}
Parse $Q$ to identify key parameters\;
Generate optimized search query $\tilde{q}$\;
% Construct metadata $M$ \;
Convert $\tilde{q}$ to vector $v_{\tilde{q}}$\;
Compute similarity scores between $v_{\tilde{q}}$ and vectors in the database\;
Retrieve top-$n$ similar contexts $C = \{c_1, c_2, \dots, c_n\}$\;

Generate expanded query set $Q_e = \{\hat{q}_1, \hat{q}_2, \dots, \hat{q}_k\}$\ using RAG based on $\tilde{q}$ and $C$\;

Generate a set of relevant keywords $K = \{k_1, k_2, ..., k_n\}$ using LLM;

\Return $\tilde{q}$, $Q_e$, $M$\, $K$;
\end{algorithm}

\vspace*{10pt}

\begin{algorithm}[H]
\caption{PubMed Search}
\label{alg:pubmed_search}
\LinesNumbered
\KwIn{User query $Q$ and keywords $K = \{k_1, k_2, ..., k_n\}$}
\KwOut{Result $P$ from PubMed}
$n \leftarrow \text{length}(K)$\;
\For{$i \leftarrow n$ \KwTo $L$}{
   $C \leftarrow$ combinations of size $i$ from $K$\;
   \For{each combo $c \in C$}{
      $q \leftarrow$ construct query using keywords $c$\;
      Send $q$ to PubMed API\;

      \If{result found}{
         $P \leftarrow$ result\;
         \Return $P$\;
      }
   }
}
\Return \text{null}\;
\end{algorithm}
\end{minipage}
\hfill
\begin{minipage}[t]{0.48\textwidth}
\begin{algorithm}[H]
\caption{Resource Identification}
\label{alg:database_search}
\LinesNumbered
\KwIn{Retrieval query $\tilde{q}$, Expanded query set $Q_e$, Keywords $K$}
\KwOut{Identified resources $\mathcal{R}$}
\BlankLine
\For{each expanded query $\hat{q}_i$ in $Q_e$}{
    Convert $\hat{q}_i$ to vector representation $v^{q_i}$\;
    Compute similarity scores between $v^{q_i}$ and document vectors $v^{d_j}$ in the database $\Delta$\;
    Retrieve top-$n$ documents $D_i = \{ d_{i1}, d_{i2}, \dots, d_{in} \}$ based on similarity scores\;
}

Merge all retrieved document sets $D_i$ into a single collection $D$\;
Merge $P$ = PubMedSearch($Q$, $K$) with $D$\;
Remove duplicates from $D$\;
Convert retrieval query $\tilde{q}$ to vector representation $v^{\tilde{q}}$\;
\For{each document $d_j$ in $D$}{
    Compute final similarity $\text{sim}_{\text{final}}(v^{\tilde{q}}, v^{d_j}) = \text{cosine\_similarity}(v^{\tilde{q}}, v^{d_j})$\;
}
Rank documents in $D$ based on $\text{sim}_{\text{final}}(v^{\tilde{q}}, v^{d_j})$\;
Get top-ranked documents as $\Delta$\;
Initialize an empty resource collection $\mathcal{R}$\;
Define a structured response schema for resource metadata\;
Use RAG with structured prompts to identify required data sources from $\Delta$\;
Extract fields such as retrieval query, data source name, data link, and paper title\;
Add identified resources to $\mathcal{R}$\;
\Return $\mathcal{R}$\;
\end{algorithm}
\end{minipage}
\end{table}

\subsection{Accessibility Module} 
This module ensures the accessibility of relevant biological databases employing a dynamic wrapper as represented in Alg. \ref{alg:enhanced_smart_wrapper}. It consists of two key functions: the Database Inspector and the Wrapper Generator.

The Database Inspector is responsible for web scraping and using information extraction techniques to locate data paths. The system extracts the access link from the metadata, leading to the queried database's source. It uses a semantic matching function to identify and filter associated data paths that use context learning for matching. The wrapper generator is responsible for data extraction from filtered data paths for the source.

The algorithm begins by identifying all relevant paths associated with the main database access link provided in the top-ranked source. These links (L) represent potential data access points, including downloadable files, HTML pages, or web forms. By employing contextual learning and semantic filtering, the algorithm narrows down these paths to a subset of filtered links $L_f$ most relevant to the user query. The next step handles the diverse nature of data sources by categorizing and processing links based on their content type.

The data access strategy of this module is based on a classification scheme that takes into account the various ways data can be stored. This approach is designed to overcome the inherent disparities between architecture and implementation that are seen in various data storage systems. In this approach, we adopt a hierarchical approach where we progressively explore different options to retrieve data, analogous to stepping back to a lower level of granularity when faced with a challenging scenario.
% We come up with an access strategy that uses a `backoff' plan. 
% Three primary categories delineate the access options.
% \begin{itemize}
%     \item \textbf{Downloadable Elements}: These comprise information that can be accessed by direct links, representing a straightforward retrieval process.
%     \item \textbf{Embedded Data Elements}: Data is stored within the page, offering a structured format for extraction.
%     \item \textbf{User Search Forms}: Information retrieved using user-initiated searches; in order to obtain the required data, particular parameters must be entered. This method may involve navigating through search forms with a search parameter.
% \end{itemize}
Finding the most accessible data options is given priority in the access strategy. Finding downloadable elements is the first priority, and then direct HTML table elements are sought for. If these choices are not available, the focus switches to user search forms, which may lead to data falling within the first two categories. we can represent the access strategy as a sequential process where we attempt different methods in a predefined order until successful retrieval of the desired data. 
% Let's denote the success of each approach as follows.
% \begin{quote}
% $S_d$: Success of finding Downloadable Elements \\
% $S_t$: Success of finding HTML Table Elements \\
% $S_f$: Success of finding User Search Forms \\
% \end{quote}
% The access strategy can then be formulated as follows.
% \begin{equation}
% \text{\small Access Strategy} = \\
% \begin{cases}
% \begin{aligned}
% &\text{\small Downloadable Elements} && \text{\small if } S_d = \text{\small True} \\
% &\text{\small HTML Table Elements} && \text{\small if } S_d = \text{\small False } \text{\small and } S_t = \text{\small True} \\
% &\text{\small User Search Forms} && \text{\small if } S_d = \text{\small False, } S_t = \text{\small False } \text{\small and } S_f = \text{\small True} \\
% &\text{\small Data Not Found} && \text{\small if } S_d = \text{\small False, } S_t = \text{\small False } \text{\small and } S_f = \text{\small False}
% \end{aligned}
% \end{cases}
% \end{equation}

% As shown in Algorithm \ref{alg:data_accessor}, w
We explore different methods of data access, progressing to the next approach only if the previous one fails. 
In this approach, the user search form processor occurs in three discrete steps:
\begin{itemize}
    \item \textbf{Form Extraction:} Complete extraction of every form on the page, together with the identification of associated buttons and inputs and their corresponding addresses.

    \item \textbf{Form Data Processor:} Leveraging Language Model (LLM) to interpret user queries and form extraction data, generating search parameters inclusive of addresses for inputs and corresponding values derived from the user query.

    \item \textbf{Form Executor:} Selenium is used to execute produced search parameters, which makes the search process easier.
\end{itemize}

Links containing downloadable files (e.g., CSV, TSV, Excel) are directly appended to the dataset list E for further ranking and selection. For links with structured tables in HTML format, the algorithm parses and converts the data into structured formats like JSON or Pandas DataFrames.

Links containing web forms are analyzed to extract form elements and input fields. A schema for form submission is generated based on the user query, and the form is programmatically submitted to access the resulting page with the search form processor.
If tables are present in the response, they are processed and appended to the dataset. Otherwise, LLM-assisted analysis dynamically generates structured data.
If multiple downloadable file links are available, they are ranked using a vector-based semantic similarity approach. The most relevant link is selected and processed into the final dataset $D_f$. If no data is successfully extracted, the source is flagged as unsuitable for automated processing. This feedback can be used to refine the system or explore alternative data access strategies.

Filtering algorithms based on metadata tags associate the correct data pieces from noisy web pages. Alg. \ref{alg:main_flow} shows the overall system and how the modules interact with each other.

% \begin{algorithm}
% \caption{Enhanced Smart Wrapper}
% \label{alg:enhanced_smart_wrapper}
% \LinesNumbered
% \KwIn{Top-ranked source $\Delta$}
% \KwOut{Generated dataset $D_f$}
% Identify associated data paths $L$ from $\Delta$\;
% Filter paths based on contextual learning to obtain $L_d$ \;
% \textbf{Process Filtered Links:}
% \Indp
%     Extract data $D_f$ <- Data Accessor($L_d$);
% \Indm \\
% \Return dataset $D_f$\;
% \end{algorithm}

\begin{algorithm}
\caption{Enhanced Smart Wrapper}
\label{alg:enhanced_smart_wrapper}
\LinesNumbered
\KwIn{Top-ranked source $\Delta$ (retrieved from the Findability module)}
\KwOut{Generated dataset $D_f$}
\textbf{Initialization:} $D_f \gets \emptyset$, $L \gets \emptyset$, $L_f \gets \emptyset$, $E \gets \emptyset$ \;

\textbf{Step 1: Identify Data Access Paths} \;
\Indp
    Extract associated links $L$ from the main database access link in $\Delta$\;
    Filter $L$ based on relevance using contextual learning to obtain $L_f$\;
\Indm

\textbf{Step 2: Analyze Filtered Links} \;
\Indp
    \ForEach{link $l \in L_f$}{
        \If{link contains downloadable file}{
            Append $l$ to $E$ (downloadable file links)\;
        }
        \ElseIf{link contains HTML table}{
            Parse the table and convert it into a structured format;
            Append the parsed table to $D_f$\;
        }
        \ElseIf{link contains web forms}{
            Identify the list of forms and their input fields\;
            Generate a form submission schema based on the user query\;
            Submit the form and extract the resulting page content\;
            \If{result contains tables}{
                Parse the table and append to $D_f$\;
            }
            \Else{
                Generate data dynamically using LLM-assisted analysis of the page\;
                Append generated data to $D_f$\;
            }
        }
        \Else{
            Flag $l$ as inaccessible or irrelevant\;
        }
    }
\Indm

\textbf{Step 3: Select the Final Data Source} \;
\Indp
    \If{$E \neq \emptyset$}{
        Perform vector similarity-based ranking on $E$ using query semantics\;
        Select the most relevant downloadable link from $E$ and append to $D_f$\;
    }
    \ElseIf{$D_f = \emptyset$}{
        Flag the source as unsuitable for data extraction\;
    }
\Indm

\Return dataset $D_f$\;
\end{algorithm}

\subsection{Interoperability Module}
The Interoperability Module ensures the integration, standardization, and processing of diverse datasets retrieved from multiple sources. While the Findability and Accessibility modules focus on identifying and retrieving data, the Interoperability Module addresses the challenges of integrating heterogeneous resources into a unified and usable format. The module is composed of two components:.

\begin{algorithm}
\caption{Main Flow}
\label{alg:main_flow}
\LinesNumbered
\KwIn{Original query $Q$}
\KwOut{Generated dataset list $\mathcal{D}_f$}
$Qs, M \leftarrow \text{QueryProcessor}(Q)$\;
$\mathcal{R} \leftarrow \text{ResourceIdentification}(Qs, M)$\;

Initialize an empty dataset list $\mathcal{D}_f$\;

\ForEach{Resource $R_i \in \mathcal{R}$}{
    $D_{f_i} \leftarrow \text{SmartWrapper}(R_i, M)$\; 
    Append $D_{f_i}$ to $\mathcal{D}_f$\;
}

\Return $\mathcal{D}_f$\;
\end{algorithm}

\subsubsection{\textbf{Process Description Generation:}}
A fundamental component of the framework is the generation of structured process descriptions, which capture the step-by-step methodology for programmatically accessing data from biological data sources on the web. These descriptions act as the foundation for automated query execution within the BioFlow system \cite{MouJ20}. Process descriptions, inspired by Needle language \cite{JamilKG2024}, enable the framework to dynamically generate executable workflows and structured queries for various data resources without manual intervention. They typically contain the following components:
\begin{itemize}
    \item A unique name that identifies the process, such as MiKDB or UniProtAccess
    \item Specifies the URL of the resource and the interaction type, e.g., access browser to indicate that the data is retrieved via a web interface.

    \item The input filters required to interact with the resource. For example, UniProtAccess requires a ProteinName string to specify the protein of interest.

    \item Defines the structure of the returned data. For instance, MiKDB outputs a table with columns such as GeneSymbol (primary key), ChrLoc (chromosomal location). UniProtAccess returns a table with Entry,	Entry Name, Protein Names, Gene Names, Organism and Length.
\end{itemize} 
This structured representation captures how to query by specifying the URL, the input form fields to fill, and the output data format. Needle can take such process descriptions and automatically construct executable queries against the described resources without manual programming. 
% \begin{quote}
% % process identifier\\
% {\sf create process} {\em MikDB}

% % access protocol\\
% \hspace*{5mm} {\sf at} {\em http://mik.bicnirrh.res.in/searchbox2.html}\\
% \hspace*{5mm} {\sf access webform} \\
% \hspace*{5mm} {\sf 'inputs': [{'xpath': //[@id="left"]/form/p/input', 'value': 'teratozospermia'}], 'submit button': //[@id="left"]/form/p/input[2]', 'other button': []}

% % output table scheme\\
% \hspace*{5mm} {\sf returns table} (\\
% \hspace*{5mm} {\em Gene Symbol} {\sf str},\\
% \hspace*{5mm} {\em Gene Name} {\sf str}\\
% );
% \end{quote}

\begin{table}[htbp!]
\centering
\caption{Details of BioFlow Process Description}
\label{tab:bioflow_process}
\begin{tabularx}{\textwidth}{@{}p{0.48\textwidth} p{0.48\textwidth}@{}}
\toprule
\textbf{MIKDB:} &
\textbf{UniProt:} \\
\midrule

% Left column
\begin{tabular}{@{}l@{}}
{\sf create process} \emph{MiKDB}\\
\quad {\sf at} \emph{http://mik.bicnirrh.res.in/mip.php}\\
\quad {\sf access browser}\\
\quad {\sf postfix} \emph{/mip.php/}\\
\quad {\sf accepts filter} (\\
\quad\quad \emph{Phenotype} {\sf String} )\\
\quad {\sf returns table} (\\
\quad\quad \emph{Symbol} {\sf GeneSymbol primary key},\\
\quad\quad \emph{ChrLoc} {\sf string},\\
\quad\quad \emph{Disease} {\sf string});
\end{tabular}
&

% Right column
\begin{tabular}{@{}l@{}}
{\sf create process} \emph{UniProtAccess}\\
\quad {\sf at} \emph{https://www.uniprot.org/}\\
\quad {\sf access browser}\\
\quad {\sf accepts filter} (\\
\quad\quad \emph{ProteinName} {\sf String} )\\
\quad {\sf returns table} (\\
\quad\quad \emph{Entry} {\sf string primary key},\\
\quad\quad \emph{EntryName} {\sf string},\\
\quad\quad \emph{ProteinNames} {\sf string},\\
\quad\quad \emph{GeneNames} {\sf string},\\
\quad\quad \emph{Organism} {\sf string},\\
\quad\quad \emph{Length} {\sf int});
\end{tabular}
\\

\bottomrule
\end{tabularx}
\end{table}

The generated process descriptions are stored in a database. This enables an efficient data retrieval mechanism for recurring queries. When a user's query matches a data source for which FAIRBridge already has a stored process description, it can directly execute that process description to generate the required data on-the-fly. This bypasses the need to go through the entire workflow of semantic search, smart wrapping, and data extraction again. By reusing stored process descriptions for known data paths, FAIRBridge saves time and computational resources compared to handling each query from scratch. This improves the overall system efficiency, especially for frequently accessed biological databases. In cases where no matching process description exists for a user's query, the system reverts to its full workflow - retrieving relevant sources, extracting datasets via web automation, and generating new process descriptions. 

\subsubsection{\textbf{Generation and Execution of BioFlow Queries:}}

The BioFlow query processing module is responsible for converting user queries into structured statements that can be executed across diverse data sources. This process ensures robustness, consistency, and precision by binding user requirements to predefined workflows and abstracting data source complexities. This step builds upon the prior query analysis phase, where the Findability Module has already identified the required data sources and retrieved their corresponding process descriptions from the Process Description Knowledgebase. By leveraging this preprocessed information, the BioFlow query processing module focuses on translating the query into a structured language. The main construction of BioFlow uses is the extract clause using from submit which is a formalization of the SQL’s select from where statement enriched with data integration features. The underlying model is truly relational, and the goal is to have a uniform view of any data as relations. Basically, the idea is to parameterize a function with other functions (matcher, filler, and wrapper), not just values, and the relational operator that abstracts extract statement is the transformation of relational operators. The basic structure of a extract statement is as outlined below:

\begin{quote}
{\sf extract} {\em Attribute List}\\
    {\sf using matcher} $\mu$ {\sf filler} $\phi$ {\sf wrapper} $\omega$\\
    {\sf from} $\varphi$\\
    {\sf submit} $r$
\end{quote}

In the above statement, the `Attribute List` returns a structured data table, and the input argument to the `extract` statement consists of the rows of values in the table \( r \). These values are passed as parameters to the optional list of functions: \( \mu \) (a schema matcher for resolving schema heterogeneity), \( \phi \) (a form-filling function for deep web form filling), and \( \omega \) (a wrapper to identify and extract table structures).

The framework begins by interpreting the natural language query provided by the user. Employing GPT-4o leveraged structured query generation, the system maps the query requirements into a BioFlow-compliant structure. 
% The Query Processor ensures that all required data sources are identified, and any existing process descriptions are retrieved, streamlining this translation process. 

Process descriptions serve as reusable workflows for accessing and extracting data from each source. If the required process descriptions are present in the Process Description Knowledgebase, the Accessibility Module incorporates them into the BioFlow query directly. This eliminates the need for redundant operations, such as regenerating wrappers or reanalyzing schemas.

If process descriptions are missing, the framework triggers the Accessibility Module to generate new process descriptions, including workflows for accessing the data source (e.g., web scraping, HTML table parsing, or form submission). These are then stored for future use and incorporated into the current query.

Using the results of the query analysis and the retrieved process descriptions, the system generates a BioFlow query that orchestrates data extraction and integration. For the example query $Q$, the resulting BioFlow statement is as follows:
\begin{quote}
% query to fetch data
{\sf select} {\em GeneSymbol, ProteinID, InfertilityData} \\
{\sf from} ( \\
\hspace*{5mm} {\sf with uniprot as (} \\
\hspace*{10mm} {\sf extract} {\em GeneSymbol, ProteinID} \\
\hspace*{10mm} {\sf using matcher} {\em S-match} \\
\hspace*{10mm} {\sf wrapper} {\em Web-Prospector} \\
\hspace*{10mm} {\sf from} {\em https://www.uniprot.org} \\
\hspace*{10mm} {\sf submit} {\em uniprot} \\
\hspace*{5mm} ), \\ 
\hspace*{5mm} {\sf mikdb as (} \\
\hspace*{10mm} {\sf extract} {\em InfertilityData} \\
\hspace*{10mm} {\sf using matcher} {\em S-match} \\
\hspace*{10mm} {\sf wrapper} {\em Web-Prospector} \\
\hspace*{10mm} {\sf from} {\em http://mik.bicnirrh.res.in/} \\
\hspace*{10mm} {\sf submit} {\em mikdb} \\
\hspace*{5mm} ) \\
) \\
{\sf where} {\em GeneSymbol = 'H2A histone'}; \\
\end{quote}

The BioFlow query further incorporates post-processing capabilities, ensuring that the data retrieved from multiple sources is aligned with user specifications and interoperable for downstream applications. Post-processing operations addresses tasks such as data filtering, merging, and transformation. This is particularly important for queries involving multiple data sources, where interoperability is a challenge due to potential differences in data formats, schemas, or access protocols.

For instance, in the example query $Q$, the BioFlow statement includes the merging of data from UniProt and the Male Infertility Knowledgebase (MiKDB). The extracted attributes from each source are combined into a unified dataset by performing a join operation. This ensures that the user receives a cohesive dataset that meets their requirements.

In this process, the extract clause retrieves data from individual sources, while the select clause performs post-processing operations to structure and integrate the data. The integration is facilitated by BioFlow's ability to express relational operations directly in its declarative syntax. 

The integration capabilities of BioFlow directly contribute to the FAIR principles, particularly interoperability. By structuring data into relational tables and executing join operations, BioFlow eliminates barriers posed by heterogeneity in data representation.

\section{Resource Discovery Process}
We now use the query $Q$ to illustrate how FAIRBridge uses the model to help find information of interest from online deep web databases.

\eat{
\begin{figure}[ht!]
\centering
\subfigure[Vector Database Search Result.]{
    \fbox{
    \includegraphics[width=.45\textwidth,keepaspectratio]{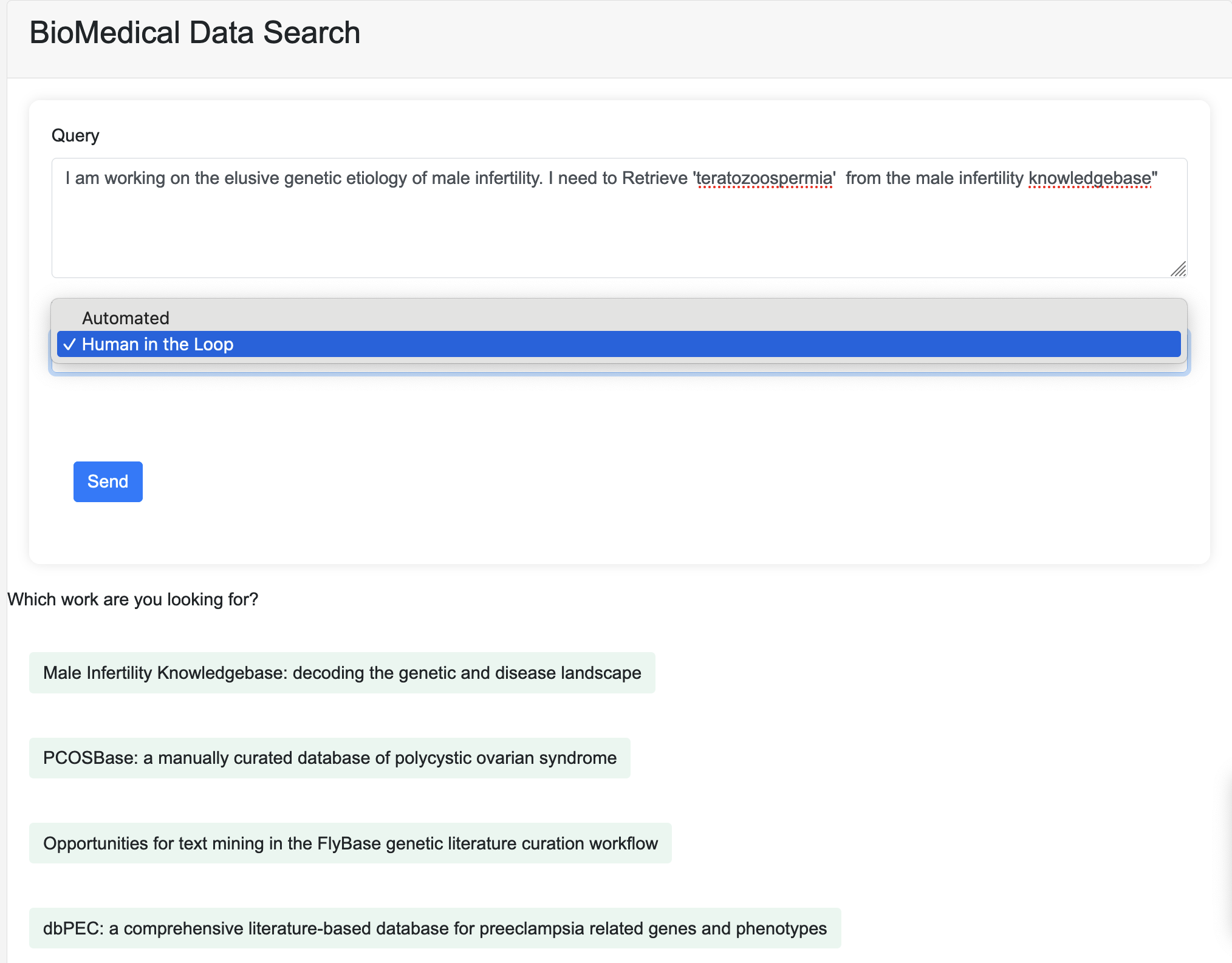}
    \label{fig:f2}
}}
%\hspace*{.75cm}
\subfigure[Abstract Details.]{
 \fbox{
 \includegraphics[width=0.45\textwidth,keepaspectratio]{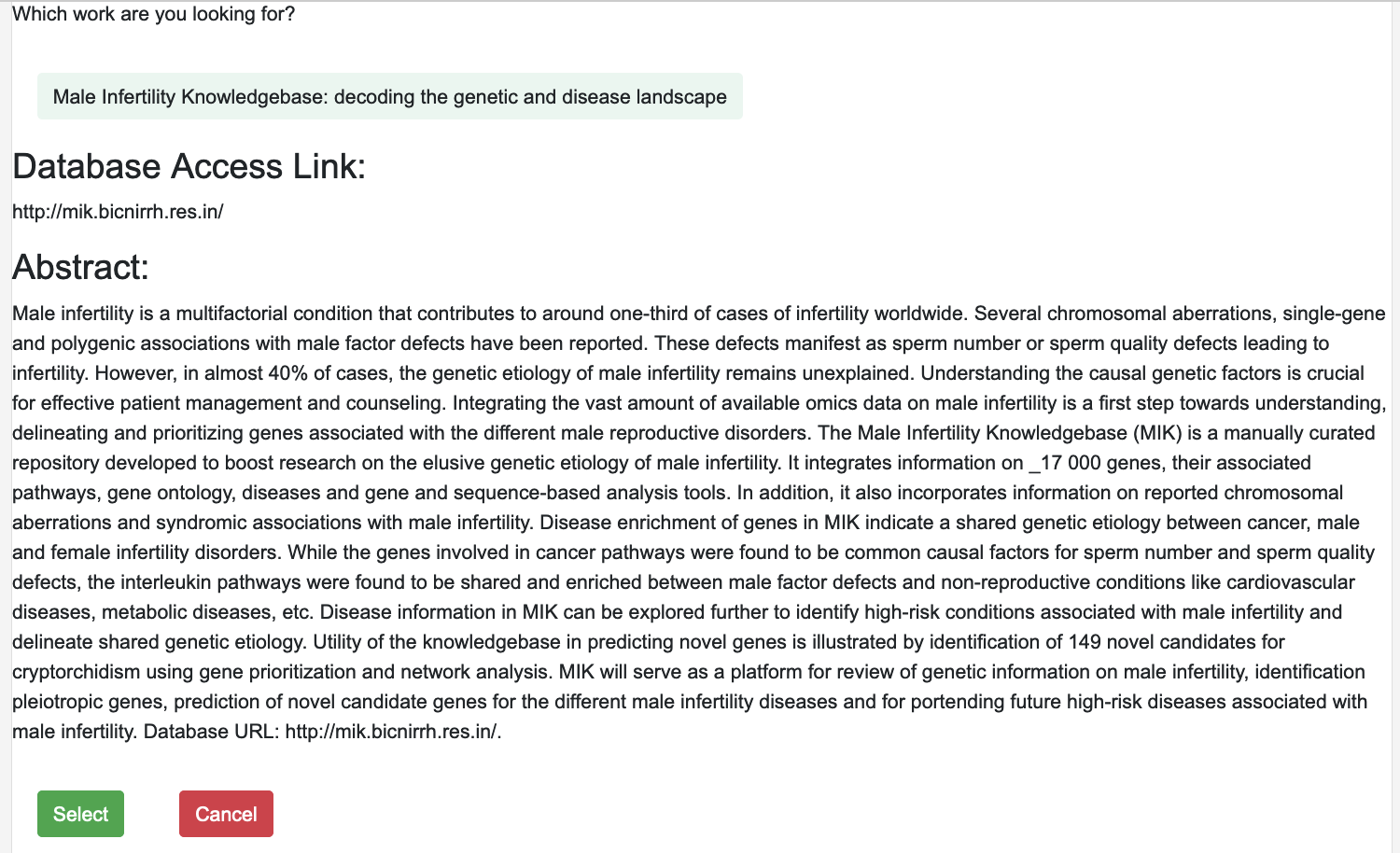}
  \label{fig:f3}
}}
\caption{Resource Discovery Process in FAIRBridge.}
\label{fig:process}
\end{figure}
}

\begin{figure}[htb!]
\centering
    \fbox{
\includegraphics[width=.9\textwidth,keepaspectratio]{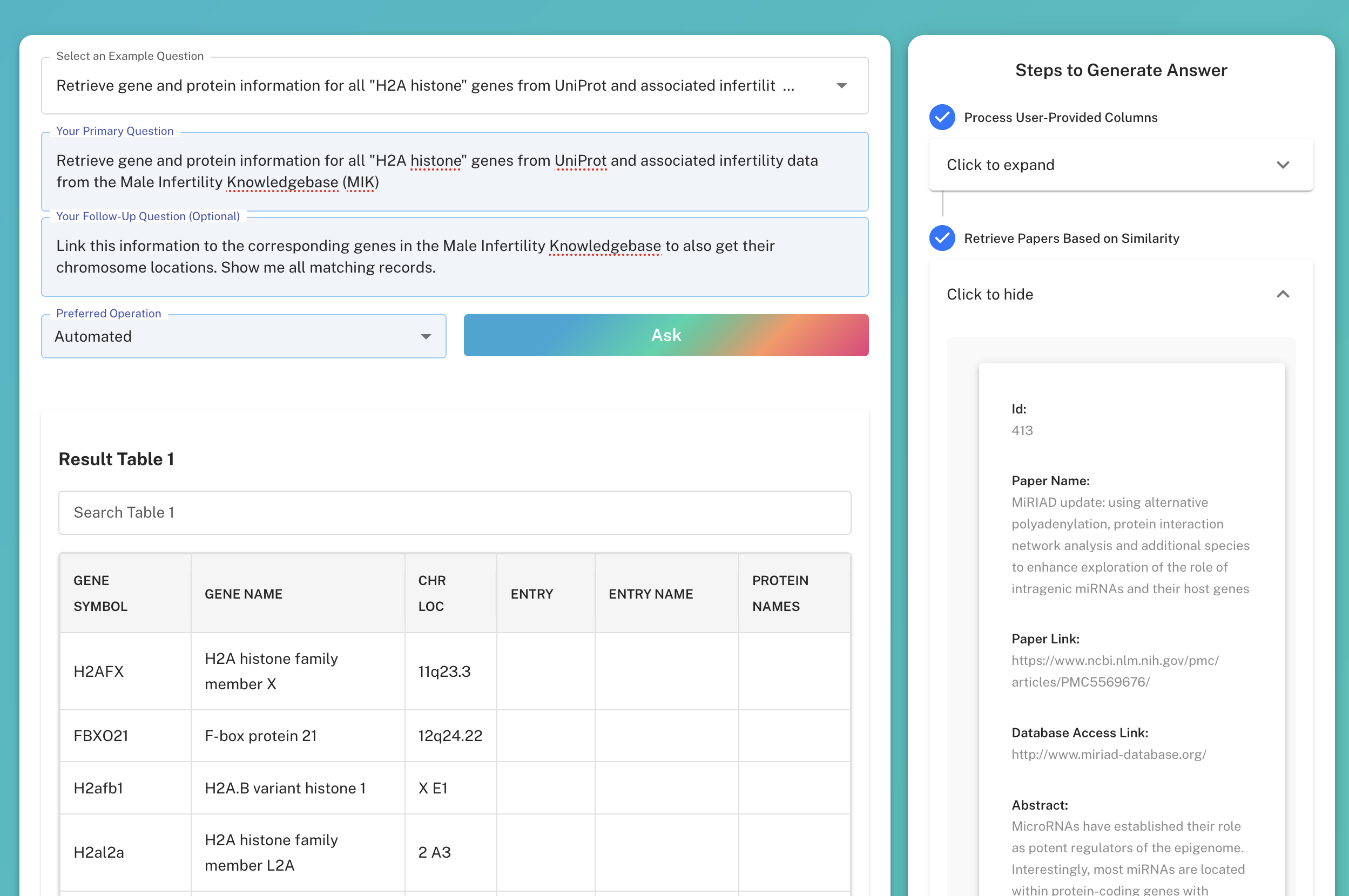}
    % \label{fig:f2}
}
\caption{ Interoperable Data from Different Sources. The user provides a query and optional follow-up for post-processing. The right panel displays stepwise logs, while the bottom presents the final merged dataset.}
\label{fig:query_multiple_post_process}
\end{figure}

% \begin{figure}[ht!]
% \centering
%     \fbox{
%     \includegraphics[width=.6\textwidth,keepaspectratio]{from_Kallol_01/multi_human_1.png}
%     % \label{fig:f2}
% }
% \caption{Resource Discovery Process in FAIRBridge: Vector Database Search Result in human-in-the-loop.}
% \label{fig:query_human_1}
% \end{figure}

We use a simple user interface to accept users' data request in natural English to initiate deep web information search. This front-end interface offers a choice of {\em automated search} or {\em human-in-the-loop search}. Once the query $Q$ is entered and a search selection is made, the process starts. 
Figure \ref{fig:fe2} in Sec \ref{tour} shows the results page of the automated approach. Using the example query $Q$, we can see the resulting dataset. The right panel displays the intermediate steps used to generate the final result. Users can pause the execution, select a different option at any intermediate step, and run the process again. This flexibility allows users to explore different query configurations.

Figure \ref{fig:query_multiple_post_process} demonstrates the post-processing capabilities of FAIRBridge, which ensures data interoperability. Users can add a follow-up query to further refine the data. In this figure, an additional follow-up query is used alongside the example query $Q$, merging the two datasets with a conditional join operation. Columns from both datasets appear in the final merged result.

\begin{figure}[htb!]
\centering
\includegraphics[width=.8\textwidth,keepaspectratio]{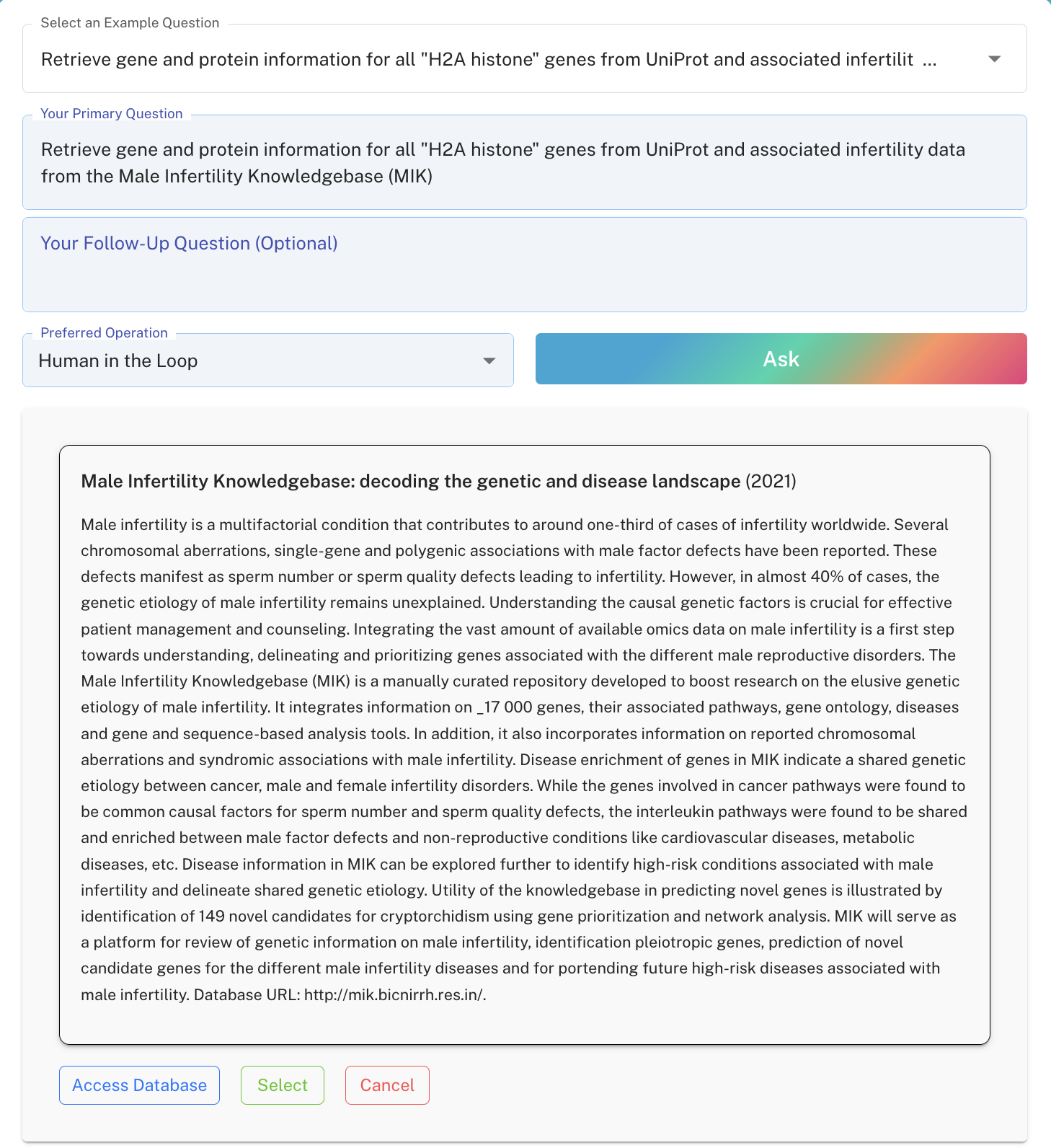}
\caption{Search Result in Human-in-the-loop Resource Discovery Process. The most relevant papers retrieved from PubMed and the vector database based on the user query are presented. Users can review the summaries of suggested resources and select the most appropriate one using the ``Access Database'' option for further processing.  }
\label{fig:multi_human}
\end{figure}

\begin{figure}[htb!]
\centering
\includegraphics[width=.7\textwidth,keepaspectratio]{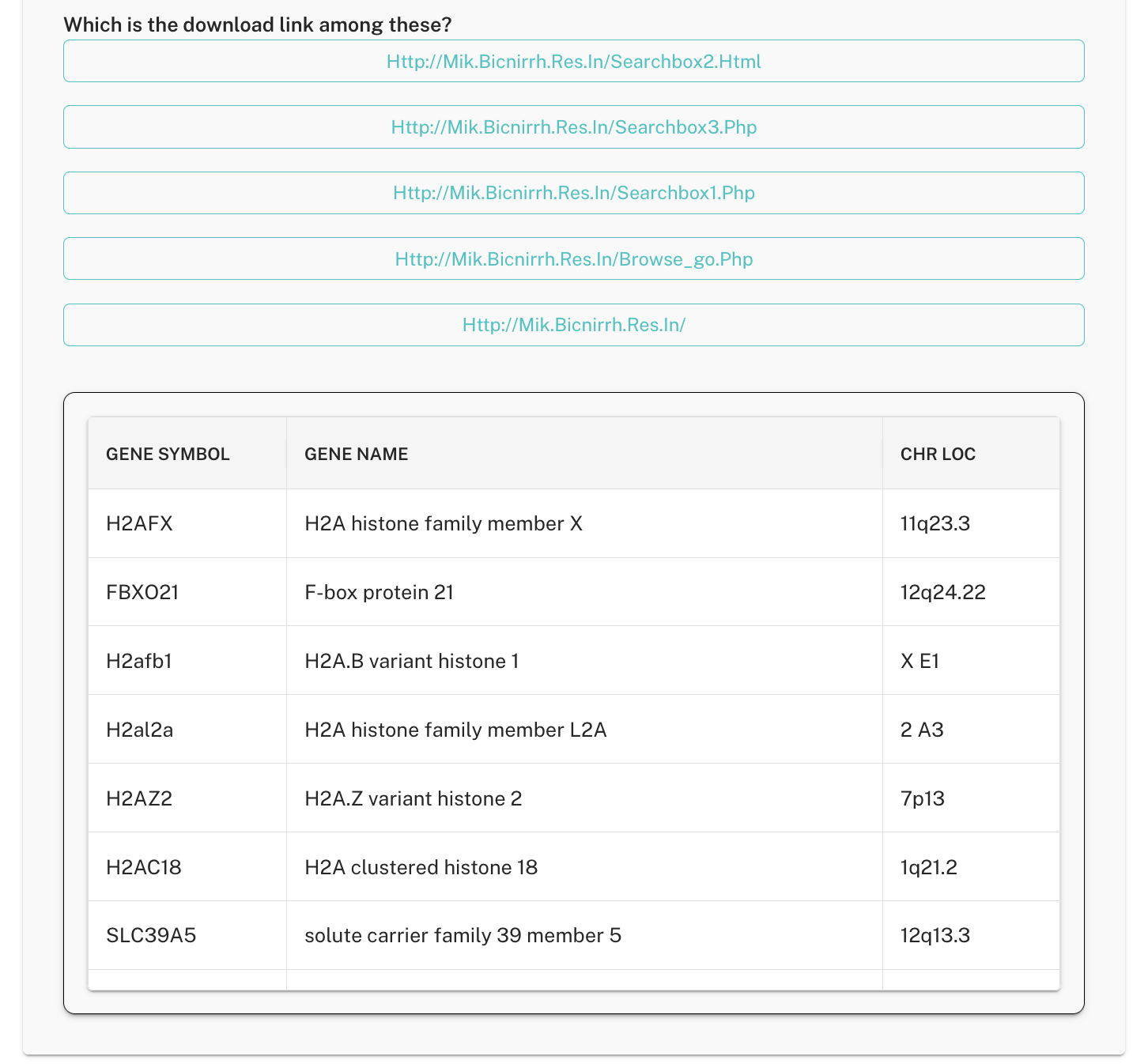}
\caption{Intermediate Stage in Human-in-the-loop Resource Discovery Process. The accessibility module's wrapper identifies and presents potential access links for the dataset. Users can select the most appropriate link, which triggers the interoperability module to process the selected resource and display the retrieved data in the table below.}
\label{fig:multiple_human_2}
\end{figure}

Fig. \ref{fig:multi_human} and Fig. \ref{fig:multiple_human_2} illustrate the subsequent stages in the human-in-the-loop approach. After the database search is conducted, the tool returns the top 4 ranked sources. The user is then prompted to choose a specific paper from which the database should be displayed. This interactive step enables users to exert control over the data selection process, ensuring alignment with their research objectives.

\eat{
Fig. \ref{fig:f3} provides a detailed view of the paper selected by the user. It encompasses vital information, including the paper name, abstract, and database access link. This level of granularity in the presentation ensures that users have access to comprehensive data from the chosen source, facilitating a deeper understanding of the research material.
}

In Fig. \ref{fig:multiple_human_2}, users are presented with the most probable download links in the {\em human-in-the-loop search}. The system prompts the user to select the link that is believed to contain the dataset of interest. This step streamlines the data acquisition process, enabling users to pinpoint the most relevant dataset link with confidence.

% We need to change this figure

\section{Experimental Setup and Evaluation}

This section explores the approaches used to assess the developed tool's findability, which facilitates data display based on user queries. 
% The evaluation introduces measures such as findability and findability bias to quantify the accessibility of information inside a collection of documents, taking inspiration from the work on findability \cite{sinha2023findability}.

\subsection{Experiment Design}

For evaluation,  we generated $m=4$ natural language queries for each of the $n=567$ documents ($D$) from PubMed.

% using which we attempted to evaluate the findability $F_Q(D)$ and Bias. To measure findability, the following steps are taken:
% \begin{itemize}
%     \item Generate two sets of relevant queries for each document using two query generation methods.
%     \item Retrieve the documents for each query using the FAIRBridge tool and record the rank of each document in the search result.
%     \item Compute the findability score for each document using the inverse law of convenience function, which is the average of the inverse of the document rank for all relevant queries, up to a threshold rank c. For our evaluation, c is set to 4.
%     \item Compute the mean findability for the collection by averaging the findability scores of all documents.
%     \item Compute the findability bias for the collection by calculating the Gini coefficient of the findability scores of all documents.

% \end{itemize}
We use two query generation strategies for measuring evaluation:
\begin{itemize}
    \item \textbf{Human-like Queries:} These natural language queries mimic real-world search behavior. They are constructed by providing contexts of documents as prompts to a GPT 4o model. Four natural language queries per document are generated based on context.
    \item \textbf{Topic-based Queries:} These queries leverage latent semantic structures. They are derived by feeding text data into Latent Dirichlet Allocation (LDA) \cite{jelodar2019latent} to extract 4 keyword clusters per document. Terms in topics form raw, unstructured queries.
\end{itemize}

\subsection{Evaluation}

This section explores the evaluation metrics utilized to assess the retrieval of PubMed abstracts, evaluate the end-to-end system performance, and analyze FAIR metric compliance.

\subsubsection{\textbf{Findability and Bias}}
Findability measures how easily a document can be found or discovered in the IR system, while Findability Bias indicates the inequality or skewness in the distribution of findability scores across documents \cite{sinha2023findability}. Higher findability scores and lower findability bias values are desirable, as they suggest better overall discoverability of documents and a more even distribution of findability scores, respectively.
For a corpus of PubMed abstracts $d_i\in D$, $1\leq i\leq n$, $n=567$, we define the findability $f$ of each document $d_i$ as the sum total of the multiplicative inverse of the ranks at which the documents are retrieved in the context of the queries ${q}_{ij}$ as follows where $q_{ij}$ is a set of $j$ queries corresponding to document $d_i$ when $1\leq j\leq m$, $m=4$.
\begin{equation}
    f(d_i)= \sum_{i=1,j=1}^{n,m} \frac{1}{r(d_i, {q}_{ij})}
\end{equation}
The overall FAIRBridge findability of a corpus of documents $D$ in the context of queries $q$ is defined as follows:
\begin{equation}
    F_q(D)= \frac{\sum_{i=1}^n f(d_i)}{n}, \mbox{where $d_i\in D$}
\end{equation}
The Gini coefficient or bias of each document $d_i\in D$ in the corpus is calculated as the function $\Xi_Q$ below:
\begin{equation}
    \Xi_Q(f(d_i)) = \frac{{\sum_{i=1}^{n} (2i - n - 1) \times f(d_i)}} {n \times \sum_{i=1}^{n} f(d_i)}
\end{equation}

% \subsubsection{\textbf{Mean Average Precision (mAP):}}
% The mAP metric measures the overall ranking quality of the system, considering both precision (ability to retrieve relevant documents) and the ranking of those relevant documents \cite{zuva2012evaluation}. Higher mAP values indicate better performance in retrieving relevant documents and ranking them higher in the result list. The mAP metric is calculated as the mean of the Average Precision (AP) values across all queries. The AP for a single query is computed as follows:
% \begin{equation}
% AP = \sum_{k=1}^{N} \frac{P(k) \times \text{rel}(k)}{R}
% \end{equation}
% where,
% \begin{itemize}
%     \item $N$ is the number of retrieved documents
%     \item  $P(k)$ is the precision at rank $k$
%     \item  $\text{rel}(k)$ is an indicator function that is equal to - 1 if the document at rank $k$ is relevant, and 0 otherwise
%     \item  $R$ is the total number of relevant documents for the query
% \end{itemize}
% The mAP is then calculated as,
% \begin{equation}
% mAP = \frac{1}{Q} \sum_{q=1}^{Q} AP(q)
% \end{equation}
% where $Q$ is the total number of queries.

\subsubsection{\textbf{Hit Rate:}}
The Hit Rate metric measures the proportion of queries for which at least one relevant document is retrieved in the top-k results (in this case, top 4). Higher Hit Rate values suggest better performance in retrieving at least one relevant document for a given query. The Hit Rate metric measures the proportion of queries for which at least one relevant document is retrieved in the top-k results. It is calculated as:
\begin{equation}
\text{Hit Rate} = \frac{1}{N_q} \sum_{i=1}^{N_q} \mathbb{I}[\exists r \leq k, d_r \in \text{Rel}(\tilde{q_i})]
\end{equation}
where,
\begin{itemize}
    \item  $N_q$ is the total number of queries 
    \item  $k$ is the cutoff rank (in this case, $k=4$) 
    \item $\mathbb{I}[\cdot]$ is the indicator function that takes the value 1 if the condition inside the brackets is true, and 0 otherwise
    \item $r$ represents the rank of a document in the retrieved list
    \item $d_r$ is the document at rank $r$
    \item $\text{Rel}({q_i})$ is the set of relevant documents for query ${q_i}$
\end{itemize}

\subsubsection{\textbf{Mean Reciprocal Rank (MMR):}}
MMR is a metric that considers the ranking of the first relevant document for each query \cite{liu2009learning}. Higher MMR values indicate that the system is better at ranking the most relevant document higher in the result list. The MMR metric is calculated as the average of the reciprocal ranks of the first relevant document for each query. It is defined as:
\begin{equation}
MMR = \frac{1}{N_q} \sum_{i=1}^{N_q} \frac{1}{\text{rank}_({q})}
\end{equation}
where,
\begin{itemize}
    \item $N_q$ is the total number of queries
    \item  $\text{rank}_({q})$ is the rank of the first relevant document for query $({q})$.
\end{itemize}

\subsubsection{\textbf{Fairsharing Evaluation Tool:}}
Fairsharing Evaluation Tool is used to evaluate the FAIRness of the biological databases. This tool provides a robust framework for assessing the FAIRness of digital resources based on a set of Maturity Indicators (MIs) \cite{sansone2019fairsharing}.
The FAIRsharing evaluation process covers significant aspects that contribute to the overall FAIRness of a dataset. These indicators encompass crucial factors such as the presence of unique identifiers, identifier persistence, metadata structure, data accessibility mechanisms, and the utilization of knowledge representation languages and controlled vocabularies.
The evaluation process involves calculating a success rate percentage for each dataset, quantifying the extent to which it satisfies the defined set of MIs. This success rate is derived using the following evaluation metric:
\begin{equation}
\text{Success Rate (\%)} = \left( \frac{\text{Number of MIs satisfied}}{\text{Total number of applicable MIs}} \right) \times 100
\end{equation}
A higher success percentage rate indicates better adherence to the FAIR principles, as the dataset fulfills a greater number of the specified MIs. Conversely, a lower success rate signifies potential areas for improvement in terms of enhancing the dataset's FAIRness.

\subsection{Results}

% Table \ref{tab:findability_assessment_metrics} shows that the human-like queries obtained a mean findability score of 0.721, indicating moderately high search accuracy in identifying relevant documents from vector store. The Gini coefficient of 0.195 points to relatively low inequality in retrievability likelihoods across corpus.
% However, the topic-based raw term queries exhibit lower accuracy, with a mean score of 0.2538 owing to lack of context. The results are summarized in the table below.
% The bias is also higher given the increased variability in search success as denoted by the Gini score of 0.44.
% Recall measures the efficacy of the system in retrieving relevant data in the dataset. GPT 4o surpasses LDA in terms of recall, achieving a value of 0.694 compared to LDA's 0.222. Therefore, queries produced by GPT 4o are more adept in capturing a greater percentage of relevant data, demonstrating its efficacy in thorough information retrieval.

The performance of the different models was evaluated on the retrieval of source documents involving 567 test documents and 2268 queries. Each query had one relevant document, and the evaluation was conducted on the top 4 retrieved documents. Six models were used for this task, including five embedding models (mxbai-embed-large-v1, text-embedding-ada-002, bge-base-en-v1.5, all-MiniLM-L6-v2, and GIST-Embedding-v0) and one topic modeling approach (LDA). The evaluation metrics used were mean findability, findability bias, hit rate, and mean reciprocal rank (MMR).

To evaluate the impact of our query analyzer, which performs query reformatting and expansion, we tested each embedding model both with and without it. Models tested with the query processor are marked with an asterisk (*) in Table \ref{tab:evaluation}. The models tested without the query processor used the raw user queries.

As shown in the Table, embedding models tested with the query analyzer consistently outperformed their counterparts that used raw user queries across all evaluation metrics.

The text-embedding-ada-002 model achieved the highest Mean Findability of 0.846 when using the query analyzer, closely followed by GIST-Embedding-v0 with 0.844 and bge-base-en-v1.5 with 0.839. Without query analyzer, these models scored lower: text-embedding-ada-002 at 0.761, GIST-Embedding-v0 at 0.759, and bge-base-en-v1.5 at 0.755. This demonstrates that the query analyzer significantly enhances document retrieval.

For Findability Bias, the text-embedding-ada-002 model had the lowest bias at 0.145 when utilizing the query processor, indicating a more uniform distribution of findability scores. Bias scores were higher without the query processor, with text-embedding-ada-002 at 0.159 and GIST-Embedding-v0 at 0.161. This suggests that the query analyzer mitigates retrieval bias.

In terms of Hit Rate, the text-embedding-ada-002 model again performed best at 0.869, followed closely by GIST-Embedding-v0 at 0.867 when using the query analyzer. This indicates that the query analyzer improves consistency in retrieving relevant documents.

The mxbai-embed-large-v1 model achieved the highest MRR of 0.980, showing that relevant documents were ranked very high on average. Models without the query analyzer had noticeably lower MRR scores.
% The evaluation results for the different models are presented in Table \ref{tab:evaluation}. According to the table, the text-embedding-ada-002 model performed the best with a mean findability of 0.846, followed closely by the GIST-Embedding-v0 model with a score of 0.844. These two models outperformed the other embedding models and the LDA model significantly, suggesting that they are more effective in retrieving relevant documents and ranking them highly for the given queries. 

% The text-embedding-ada-002 model achieved the lowest findability bias of 0.145, indicating that it performs well in making documents easily discoverable and has a more even distribution of findability scores across documents. The GIST-Embedding-v0 model also performed well in terms of findability bias, with a score of 0.147. The text-embedding-ada-002 model had the highest Hit Rate of 0.869 , followed by the GIST-Embedding-v0 model with a Hit Rate of 0.867 . These models outperformed the others in terms of retrieving at least one relevant document in the top 4 results for a higher proportion of queries. The mxbai-embed-large-v1  model achieved the highest MMR of 0.98, closely followed by the bge-base-en-v1.5 model with an MMR of 0.974. These models performed well in ranking the most relevant document higher for the given queries.

The end-to-end system result is conducted using the text-embedding-ada-002 model, which exhibited the best performance in the source retrieval stage to retrieve PubMed abstracts. By ``end-to-end evaluation'', we refer to whether the final data is successfully retrieved based on the user query. Among the 567 data sources, a total of 268 data sources encountered errors, including 404 (Not Found) errors, 502 (Bad Gateway) errors, and timeouts. These issues were beyond the scope of our tool's control, as they were caused by external factors such as unavailable or unresponsive data servers.
Additionally, 76 datasets were not found in their respective data sources.  Datasets from 65 data sources were found to be incompatible with our system's current capabilities. These included cases where the data was not presented in any tabular format, data required authentication. As a result, we consider the remaining 158 data sources for the end-to-end evaluation.

The evaluation metrics for the end-to-end system, which retrieves and generates the final data from queries, include a Mean Findability of 0.127, a Findability Bias of 0.86, and a Hit Rate of 12.7%.
% The evaluation metrics for the end-to-end system, responsible for retrieving and generating the final data from queries, are as follows.
% \begin{quote}
%     Mean Findability:  0.127 \\
% Findability Bias: 0.86 \\
% Hit Rate: 12.7\% \\
% \end{quote}

% \begin{table}[htbp]
% \centering
% \small
% \caption{Findability Assessment Metrics}
% \label{tab:findability_assessment_metrics}
% \renewcommand{\arraystretch}{1.05}
% \begin{tabular}{|p{2cm}|p{2cm}|p{2cm}|p{1cm}|}
% \hline
% \textbf{Query Generation Method} & \textbf{Mean Findability Score} & \textbf{Gini Coefficient (Bias)} & \textbf{Recall} \\
% \hline
% GPT 4o & 0.721 & 0.195 & 0.694 \\
% LDA & 0.253 &  0.44 & 0.222 \\
% \hline
% \end{tabular}
% \end{table}

% \begin{table}[htbp]
% \centering
% \small
% \caption{Evaluation of Different Models on Document Retrieval}
% \label{tab:evaluation}
% \renewcommand{\arraystretch}{1.05}
% \begin{tabular}{|p{3cm}|p{1.5cm}|p{1.5cm}|p{1.5cm}|p{1.5cm}|p{1.5cm}|}
% \hline
% \textbf{Model} & \textbf{mAP} & \textbf{Findability} & \textbf{Bias} & \textbf{Hit Rate} & \textbf{MMR} \\
% \hline
% mxbai-embed-large-v1 & 0.7527 & 0.75 & 0.1839 & 0.7655 & 0.9833 \\
% text-embedding-ada-002 & 0.8377 & 0.8321 & 0.1216 & 0.8564 & 0.9781 \\
% SFR-Embedding-Mistral & 0.8265 & 0.8206 & 0.1327 & 0.8373 & 0.9871 \\
% all-MiniLM-L6-v2 & 0.7304 & 0.7258 & 0.1905 & 0.7607 & 0.9601 \\
% GIST-Embedding-v0 & 0.7850 & 0.7802 & 0.1615 & 0.8229 & 0.9539 \\
% LDA & 0.25810 & 0.2538 & 0.4403 & 0.3264 & 0.8333 \\
% \hline
% \end{tabular}
% \end{table}

\begin{table}[htbp!]
\centering
\small
\caption{Evaluation of Different Models on Document Retrieval}
\label{tab:evaluation}
\renewcommand{\arraystretch}{1.05}
\begin{tabular}{|p{3.5cm}|p{2.5cm}|p{2.25cm}|p{1.2cm}|p{1.2cm}|p{1.2cm}|}
\hline
\textbf{Model} & \textbf{Mean Findability}  & \textbf{Findability Bias} & \textbf{Hit Rate} & \textbf{MMR} \\
\hline
mxbai-embed-large-v1*    & 0.825       & 0.163 & 0.841    & \textbf{0.980 } \\ 
text-embedding-ada-002*  & \textbf{0.846 }     & \textbf{0.145} & \textbf{0.869}    & 0.973 \\ 
all-MiniLM-L6-v2*        & 0.795       & 0.184 & 0.836    & 0.951 \\ 
GIST-Embedding-v0*       & 0.844      & 0.147 & 0.867    & 0.972 \\ 
bge-base-en-v1.5*        & 0.839       & 0.150 & 0.862    & 0.974 \\ 
\hline

mxbai-embed-large-v1    & 0.742  & 0.179 & 0.756 & 0.882  \\
text-embedding-ada-002  & 0.761  & 0.159 & 0.782 & 0.875  \\
all-MiniLM-L6-v2        & 0.715  & 0.202 & 0.752 & 0.855  \\
GIST-Embedding-v0       & 0.759  & 0.161 & 0.780 & 0.874  \\
bge-base-en-v1.5        & 0.755  & 0.165 & 0.775 & 0.876  \\
LDA                    & 0.272       & 0.423 & 0.324    & 0.84  \\ 

\hline
\end{tabular}
\end{table}

A set of 20 maturity indicators of the fairness assessment tool is used for the evaluation of Findability, Accessibility and Interoperability. Significant aspects like unique identifiers, identifier persistence, metadata structure, data accessibility and knowledge representation languages are considered as maturity indicators. The biological datasets listed in Table \ref{tab:fair_evaluation} are sourced from `Database: The Journal of Biological Databases and Curation'. These datasets are retrieved by FAIRBridge and are used for the FAIR evaluation.

\begin{table}[htbp!]
\centering
 \small
\caption{FAIRSharing Test on Successfully Retrieved Data}
\label{tab:fair_evaluation}
\renewcommand{\arraystretch}{1.05}

% -- Four columns, no vertical lines (booktabs style) --
\begin{tabular}{p{4.2cm} p{2.7cm} p{2.1cm} p{2cm}}
\toprule
\textbf{Dataset} & \textbf{FAIRSharing Success Rate (\%)} & \textbf{FAIRBridge Retrievability} \\
\midrule

CHOmine \cite{gerstl2017chomine} & 15 & \checkmark  \\
COMPARTMENTS \cite{binder2014compartments} & 15 & \checkmark  \\
CoDNaS 2.0 \cite{monzon2016codnas} & 35 & \checkmark  \\
DGPD \cite{hu2022dgpd} & 15 & \checkmark  \\
Drug-Path \cite{zeng2015drug} & 15 & \checkmark  \\
EcoliNet \cite{kim2015ecolinet} & 15 & \checkmark  \\
Gene regulation knowledge \cite{tripathi2016gene} & 15 & \checkmark  \\
MiKDB \cite{JosephM2021} & 15 & \checkmark  \\
PharmaKoVariome \cite{kim2022pharmakovariome} & 15 & \checkmark  \\
Reactome \cite{viteri2019reactome} & 45 & \checkmark  \\
SDADB \cite{zeng2018sdadb} & 35 & \checkmark  \\
CTD \cite{davis2011curation} & 35 & \checkmark  \\
UniProt Knowledgebase \cite{magrane2011uniprot} & 15 & \checkmark \\
Covid19census \cite{zanettini2021covid19census} & 35 & \checkmark  \\
Harmonizome \cite{rouillard2016harmonizome} & 45 & \checkmark \\
AnthraxKP \cite{feng2022anthraxkp} & 15 & \checkmark  \\
Prototheca-ID \cite{dziurzynski2021prototheca} & 15 & \checkmark \\
BC-TFdb \cite{khan2021bc} & 15 & \checkmark  \\
\bottomrule
\end{tabular}
\end{table}

The test observations from the FAIR evaluation throughout the datasets show a varied landscape of FAIR compliance. FAIRSharing Metrics reveal a comparatively poor success rate for the assessed datasets. The values, ranging from 15\% to 45\%, underscore certain shortcomings or areas where the datasets could enhance their adherence to FAIR principles. Some datasets demonstrated success in certain indicators, such as the inclusion of unique identifiers, adherence to metadata protocols, and the ability to implement authentication and authorization within their resolution protocols. However, they encountered difficulties in other metrics including the persistence tests for both data and metadata identifiers. These scores are reflective of the intricacies and challenges associated with achieving comprehensive FAIRness, as defined by the FAIRSharing framework.

In contrast, the FAIRBridge tool, designed with a specific focus on machine readability, is able to retrieve all of them. The FAIRBridge tool's ability to retrieve data from datasets with lower FAIRSharing scores underscores its efficacy in handling datasets with sub-optimal FAIR scores.

\subsection{Discussion}
This research introduces a novel system to enhance the findability, accessibility, and interoperability of vast amounts of unFAIR biological data. FAIRBridge integrates advanced methods such as neural semantic search, web mining, and data scraping into an adaptable, end-to-end workflow. By formalizing methods to achieve FAIR principles, we have implemented algorithms that bridge the gap between user queries and diverse data sources. The system utilizes a rank function for document retrieval and enables user-specific data extraction through schema matching and web scraping.

Embedding models like text-embedding-ada-002 and GIST-Embedding-v0 demonstrated strong performance due to their ability to capture semantic relationships, outperforming traditional models like LDA in both findability and relevance. While mxbai-embed-large-v1 achieved the highest MMR of 0.98, its lower findability scores highlight variations in model performance across different metrics. These findings underscore the effectiveness of embedding-based approaches in handling complex information retrieval tasks and their superiority in ranking relevant documents within heterogeneous datasets.

The variation in model performance across different metrics highlights the distinct strengths and limitations inherent to each approach. Some models excel at ranking highly relevant documents, while others demonstrate superior discoverability. The consistent improvements observed with the query analyzer underscore its effectiveness in bridging semantic gaps between user queries and document content. By reformatting and expanding queries, the query analyzer significantly enhances alignment and retrieval accuracy, ensuring a more efficient search process.

The qualitative analysis of IR systems shows that FAIRBridge represents a significant advancement in dataset retrieval tools, offering features that collectively are not available in existing IR systems. Its ability to process natural language queries, conduct cross-portal searches, and handle unFAIR data provides researchers with unparalleled access to biological datasets. By integrating customizable outputs and actual data presentation, FAIRBridge addresses a critical gap in current retrieval tools, delivering both data and insights in a researcher-friendly format.

While the initial stage of retrieving sources from PubMed abstracts demonstrated strong potential, challenges emerged in the final stages of data generation due to the heterogeneous nature of databases. The unique layouts, form structures, and multi-step interactions required by some sites complicate the extraction and integration processes. Although the tool effectively processes simple user search forms, it struggles with complex forms requiring multi-page submissions or intricate inputs. This limitation highlights the need for further research and development to enhance the system's flexibility and scalability in accessing deep web databases.

FAIRBridge's performance is further constrained by the current scope of its process description knowledgebase, which dictates its ability to identify and extract data elements accurately. As the knowledgebase grows with repeated use, the system will become increasingly adept at autonomously identifying and prioritizing suitable resources, even in cases where they are not explicitly mentioned in user queries. This adaptive learning capability is poised to enhance the system's robustness and its effectiveness in tackling a wider range of scientific queries.

A pivotal advancement within FAIRBridge lies in its Interoperability Module, which employs select-project-join (SPJ) queries to harmonize data from multiple sources. This module ensures the integration and usability of retrieved datasets, creating a unified data structure. However, the current system has limitations in addressing highly complex, domain-dependent queries requiring advanced reasoning or specialized domain knowledge, such as those involving intricate biological pathways or experimental conditions. Enhancing this capability remains a key focus for future development, with the goal of supporting more nuanced scientific inquiries.

Despite these limitations, FAIRBridge represents a significant step forward in addressing the challenges of unstructured and heterogeneous biological data. Unlike existing tools that merely assess FAIRness, FAIRBridge actively enhances practical findability and accessibility without altering the original data sources. This unique approach complements existing FAIR initiatives by making valuable, yet unFAIR, data more discoverable and usable. By navigating and retrieving datasets with low FAIR compliance scores, FAIRBridge not only improves data accessibility but also lays the groundwork for enhancing overall FAIRness in scientific research, healthcare, and beyond.

\section{Conclusion and Future Research}
A novel data retrieval system is presented in this study to address biological datasets' pervasive noncompliance with FAIR standards. Our tool FAIRBridge facilitates the discovery, extraction, and processing of unFAIR data sources to retrieve customizable datasets on-demand. The system architecture integrates neural semantic search, web data extraction, and automated formatting techniques for an end-to-end solution tailored to user specifications. We construct a vector database of 567 biological publications spanning 15 years as the backbone for semantic search. Natural language queries are reformatted and expanded into optimized vector representations to retrieve the most relevant papers and databases with high precision. Our smart wrapper leverages web automation and NLP to extract key information and data assets from retrieved sources. The interoperability module further filters and formats results to user-defined criteria.

Quantitative analysis indicates that the tool demonstrates strong search findability with human-like queries, achieving a mean findability score of 0.846 and a hit rate of 86.9\%. This value signifies a moderately high search accuracy in identifying relevant data sources. A comparison of assessment of biological datasets shows that FAIRBridge is able to retrieve data even when the data shows poor performance in FAIR evaluator tools.
Future improvements include enhancing the smart wrapper for complex inputs and diverse data formats, developing a specialized query processing model, and optimizing automated and human-in-the-loop processes. By focusing on Reusability, we aim to ensure that the datasets retrieved and processed by FAIRBridge can be easily repurposed and integrated into various research contexts and analytical workflows. These upgrades will increase the system's adaptability, ease of use, and efficiency, making it a versatile tool for various research needs.

\section*{Software Availability}

The first edition of FAIRBridge is available for public use at http://dblab.nkn.uidaho.edu/fairbridge/. While parts of the system are still under development, and are in flux, all functions described in this article are active.

\section*{Acknowledgment}

This research was supported in part by a National Institutes of Health IDeA grant P20GM103408, a National Science Foundation CSSI grant OAC 2410668, and a US Department of Energy grant DE-0011014.

\bibliographystyle{ACM-Reference-Format}
\bibliography{references, ref}

\end{document}